\documentclass[%
 reprint,
superscriptaddress,
 amsmath,amssymb,
 aps,
 prb,
]{revtex4-2}

\usepackage{CJK}
\usepackage{graphicx}
\usepackage{dcolumn}
\usepackage{bm}
\usepackage{float} 
\usepackage[normalem]{ulem} 

\usepackage{xcolor} 

\usepackage{hyperref}

\newcommand{\sect}{Sec.~}
\newcommand{\eqn}{Eq.~}

\newcommand{\fig}{Fig.~}

\begin{document}
\begin{CJK*}{UTF8}{bsmi}

\preprint{APS/123-QED}

\title{Anisotropic transport in ballistic bilayer graphene cavities}

\author{Florian Schoeppl}
\affiliation{University of Regensburg, Germany}

\author{Alina Mre\'nca-Kolasi\'nska}
\affiliation{AGH University of Krakow, Poland}

\author{Ming-Hao Liu (劉明豪)}
\affiliation{Department of Physics and Center of Quantum Frontiers of Research and Technology (QFort), National Cheng Kung University, Tainan 70101, Taiwan}

\author{Korbinian Schwarzmaier}
\affiliation{University of Regensburg, Germany}

\author{Klaus Richter}
\affiliation{University of Regensburg, Germany}

\author{Angelika Knothe}
\affiliation{University of Regensburg, Germany}



\date{\today}

\begin{abstract}
Closing the gap between ray tracing simulations and experimentally observed electron jetting in bilayer graphene (BLG), we study all-electronic, gate-defined BLG cavities using tight-binding simulations and semiclassical equations of motion. Such cavities offer a rich playground to investigate anisotropic electron transport due to the trigonally warped Fermi surfaces. 
{In this work, we achieve two things: First, we verify the existence of triangular modes (as predicted by classical ray tracing calculations) in the quantum solutions of closed circular BLG cavities. Then, we explore signatures of said triangular modes in transport through open BLG cavities connected to leads. We show that the triangular symmetry translates into anisotropic transport and present an optimal setup for experimental detection of the triangular modes as well as for controlled modulation of transport in preferred directions.}

\end{abstract}

\maketitle
\end{CJK*}

\section{\label{sec:Introduction} Introduction}
Recent years have seen intense scientific interest in two-dimensional atomic lattices, driven by experimental accessibility to an ever-growing range of low-dimensional materials. At the forefront of this development was the exfoliation of graphene, which has now enabled access to a variety of graphene-based van der Waals heterostructures.

One promising platform for complex 2D-material-based systems is bilayer graphene (BLG). Its tunable band gap \cite{oostingaGateinducedInsulatingState2008,pereiraTunableQuantumDots2007} allows to experimentally realise electrostatically confined nanostructures with a variety of gate-defined geometries \cite{ingla-aynesSpecularElectronFocusing2023,overwegTopologicallyNontrivialValley2018, mollerImpactCompetingEnergy2023, iwakiriGateDefinedElectronInterferometer2022}
and to explore the physically rich low-energy band structure. In the low-energy regime close to the Dirac points, the band structure of BLG exhibits trigonally warped Fermi surfaces, leading to anisotropic group-velocity distributions, cf.~\fig \ref{fig:ConductanceSetupBandstructure}.
This anisotropy leads to anisotropic transport 
\cite{peterfalviIntrabandElectronFocusing2012,rudenkoAnisotropicEffectsTwodimensional2024,zhaoInplaneAnisotropicElectronics2020} as has been observed, e.g., as distinct electron jets with scanning gate microscopy \cite{goldCoherentJettingGateDefined2021}, and in direction-dependent transverse electron focusing experiments \cite{ingla-aynesSpecularElectronFocusing2023,klanurakProbingAnisotropicFermi2024}. At even lower energies, the Fermi surface undergoes a Lifshitz transition, splitting the triangular Fermi surface into three distinct mini-valleys, altering transport properties once more \cite{shiTunableLifshitzTransitions2018,varletTunableFermiSurface2015,varletAnomalousSequenceQuantum2014,ahmedDetectingLifshitzTransitions2025}. The combination of interesting electronic properties with the experimental feasibility of creating gate-defined devices makes BLG an ideal platform for studying exotic quantum transport phenomena in realistic device geometries \cite{chakrabortiElectronWaveQuantum2024}.

{\parfillskip=0pt 
Here, we present an analysis of gate-defined, all-electronic BLG cavities using tight-binding approaches \cite{chenFourbandEffectiveSquare2024,yuanElectronicTransportDisordered2010} and semiclassical equations of motion.
 For closed cavities, our studies complement previous classical ray-tracing approaches with hard wall boundary wave matching algorithms \cite{seemannComplexDynamicsCircular2024,seemannSteeringInternalOutgoing2024,seemannGatetunableRegularChaotic2023} 
(which showed that the breaking of rotational symmetry by the anisotropic Fermi surface leads to a mixed phase space
\begin{figure}[H]
    \centering
    \includegraphics[width=0.9\linewidth]{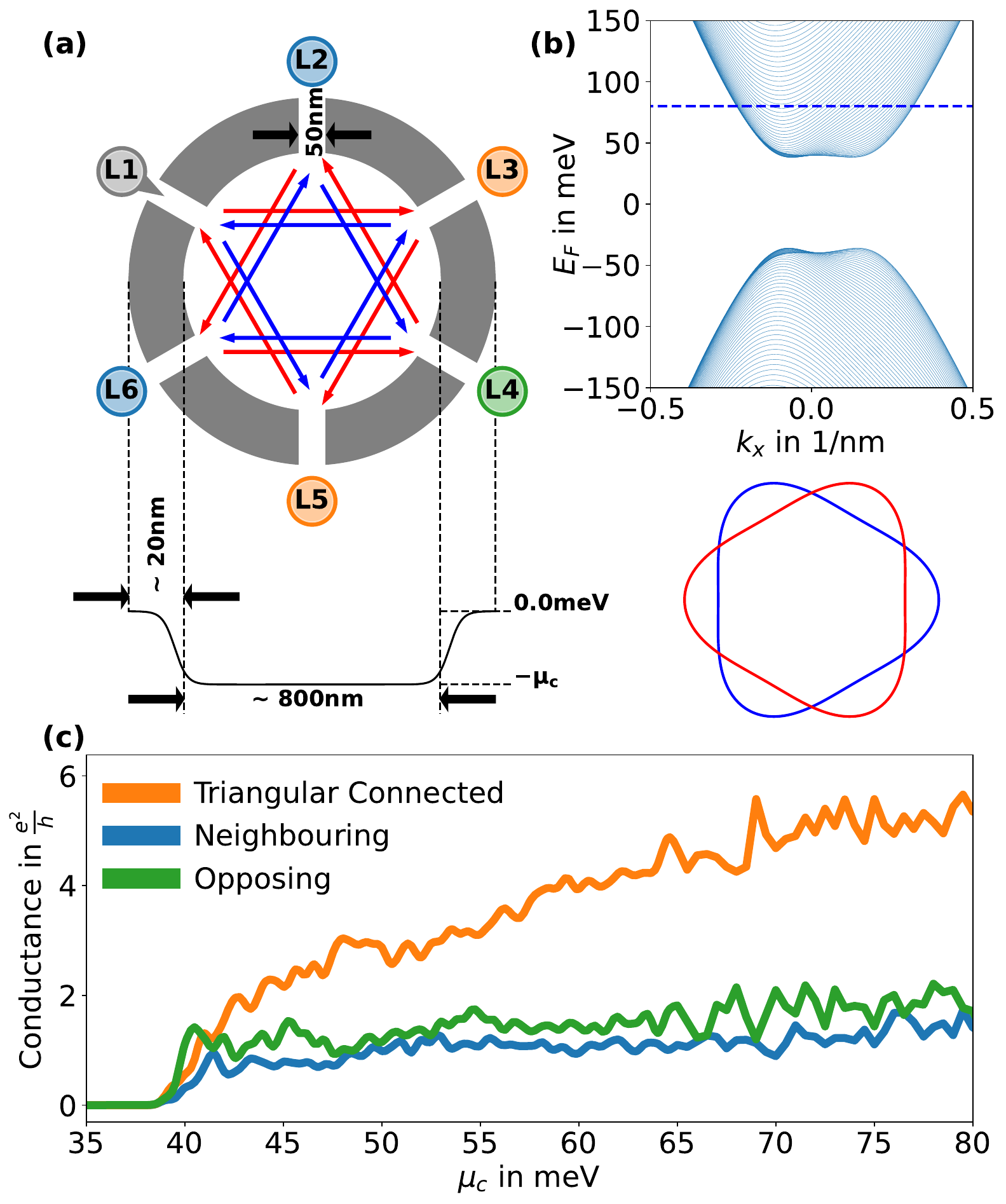}
    \caption{(a) Schematics of an electronic BLG cavity defined by a smooth, electrostatic confinement with local energy band offset $\mu_c$. (b) BLG band structure ($K^+$ valley) 
    and Fermi contours for Fermi energy $E_F = 80$ meV  (both valleys $K^\pm$). The flat sides of the triangular warped Fermi lines give rise to six dominant group velocities. (c) Conductance through the cavity for all lead combinations. Conductance between leads connectable via the dominant group velocities (red and blue arrows in (a) is strongly enhanced compared to other lead combinations.}
    \label{fig:ConductanceSetupBandstructure}
\end{figure}
\noindent  with regular and chaotic regions even in circular cavities) by including realistic, smooth boundary conditions of  electrostatic confinement. We demonstrate an enhancement of the cavity eigenstates' probability density along certain classically predicted triangular orbits. Beyond that, we study transport through open cavities, where we find distinct signatures of the cavity states' triangular structure in the anisotropic transmission properties of the cavity with and without a magnetic field, see \fig \ref{fig:ConductanceSetupBandstructure} \footnote{The conductance data in \fig \ref{fig:CompareHeatmapConductance} has been smoothed by applying a thermal broadening of 2 K for better readability}. 
Finally, based on the stable periodic orbits' shape and the system's symmetry, we demonstrate that a lead setup as displayed in \fig \ref{fig:ConductanceSetupBandstructure} allows for measuring these anisotropic transport effects independently of the lattice-to-device orientation.
In earlier work \cite{schrepferDiracFermionOptics2021}
on Dirac fermion optics in bilayer graphene cavities with smooth confinement, related features have been found in the resonant states of microdisks where, moreover, the deformation of the disks leads to directed emission of charge carriers.  
}

Microscopy tools, such as scanning tunneling microscopy, have been used previously successfully to access the rich quantum states of graphene-based billiards \cite{keski-rahkonenVariationalScarringGraphene2025,geDirectVisualizationRelativistic2024,cabosartRecurrentQuantumScars2017}
. Here, we propose using electronic transport to extract information about the confined cavity states and to obtain controlled, directional transmission through such a device, enabled by the symmetry of the underlying material's electronic structure.

\section{Quantum Solutions of a Closed Cavity}

First, we discuss the bound states of a closed, circular BLG cavity.
We model the  properties of BLG using the four-band-model \cite{mccannElectronicPropertiesBilayer2013},
\begin{align}
    \hat{H} = \left( \setlength{\arraycolsep}{2pt}\begin{array}{cccc}
	\epsilon_{A1} & v_0\pi^\dagger  & 0 & v_3\pi\\
	v_0\pi & \epsilon_{B1} & \gamma_1 & 0 \\
	0 & \gamma_1 & \epsilon_{A2} & v_0\pi^\dagger \\
	v_3 \pi^\dagger & 0 & v_0\pi & \epsilon_{B2}
    \end{array}\right), \,
     v_i = \frac{\sqrt{3} a \gamma_i}{2\hbar}, \; i\in \{0,3\},
    \label{eqn:HBLG}
\end{align}
with velocities, $v_i$,
in which $a = 0.246$ nm is the BLG lattice constant, $\gamma_0 = 3160$ meV the intralayer hopping, $\gamma_1=381$ meV the vertical interlayer hopping between dimer sites, and $\gamma_3 = 380$ meV is the skew hopping between  non-dimer sites \footnote{Note that we do not feature $\gamma_4$ in \eqn\eqref{eqn:HBLG}, as this parameter introduces electron-hole asymmetry and we focus exclusively on electron transport.}.   
On-site terms are given by,
\begin{align}
 \nonumber    \epsilon_{A1/B2} &= \mp\frac{U}{2}  + \mu(\mathbf{r}), 
    \quad\epsilon_{B1/A2} = \frac{1}{2} (\mp U + 2 \Delta') + \mu(\mathbf{r}),
\end{align}
with the electrical displacement field $U$,    the difference between dimer and non-dimer sites $\Delta' = 22$ meV, and the local potential landscape $\mu(\mathbf{r})$, with the real space position $\mathbf{r}=(x,y)$.
We model a closed, electrostatically confined, circular cavity by a spatial varying potential,
\begin{align}
    \mu(\mathbf{r}) = \frac{\mu_c}{2} \left(1+\tanh\left(\frac{4}{d}(|\mathbf{r}|-r_{c})\right)\right) -\mu_c , \label{eq:Transition Function}
\end{align}
where $\mu_c$ is the local band offset inside the cavity, $d = 20$ nm a smoothing constant, and $r_c = 400$ nm the cavity radius.
The Fermi energy outside the cavity is fixed at  0 meV inside the band gap.
In \eqn\eqref{eqn:HBLG}, we write the momentum parameters as,
\begin{align}
 \nonumber   \pi &= p_x (\xi \cos(\alpha) + i \sin(\alpha)) + p_y (-\xi \sin(\alpha)+i \cos(\alpha)) \\
    \pi^\dagger &= p_x (\xi \cos(\alpha) - i \sin(\alpha)) + p_y (-\xi \sin(\alpha)-i \cos(\alpha)),
    \label{eqn:pi}
\end{align}
where $\mathbf{p} = (p_x,p_y)$ is the momentum centered at the $K^\xi$, $\xi \in \{-1,1\}$, point.
The parameter $\alpha$ controls the rotation of the Hamiltonian in the $p_xp_y$-plane.

\begin{figure}
    \centering
    \includegraphics[width=0.9\linewidth]{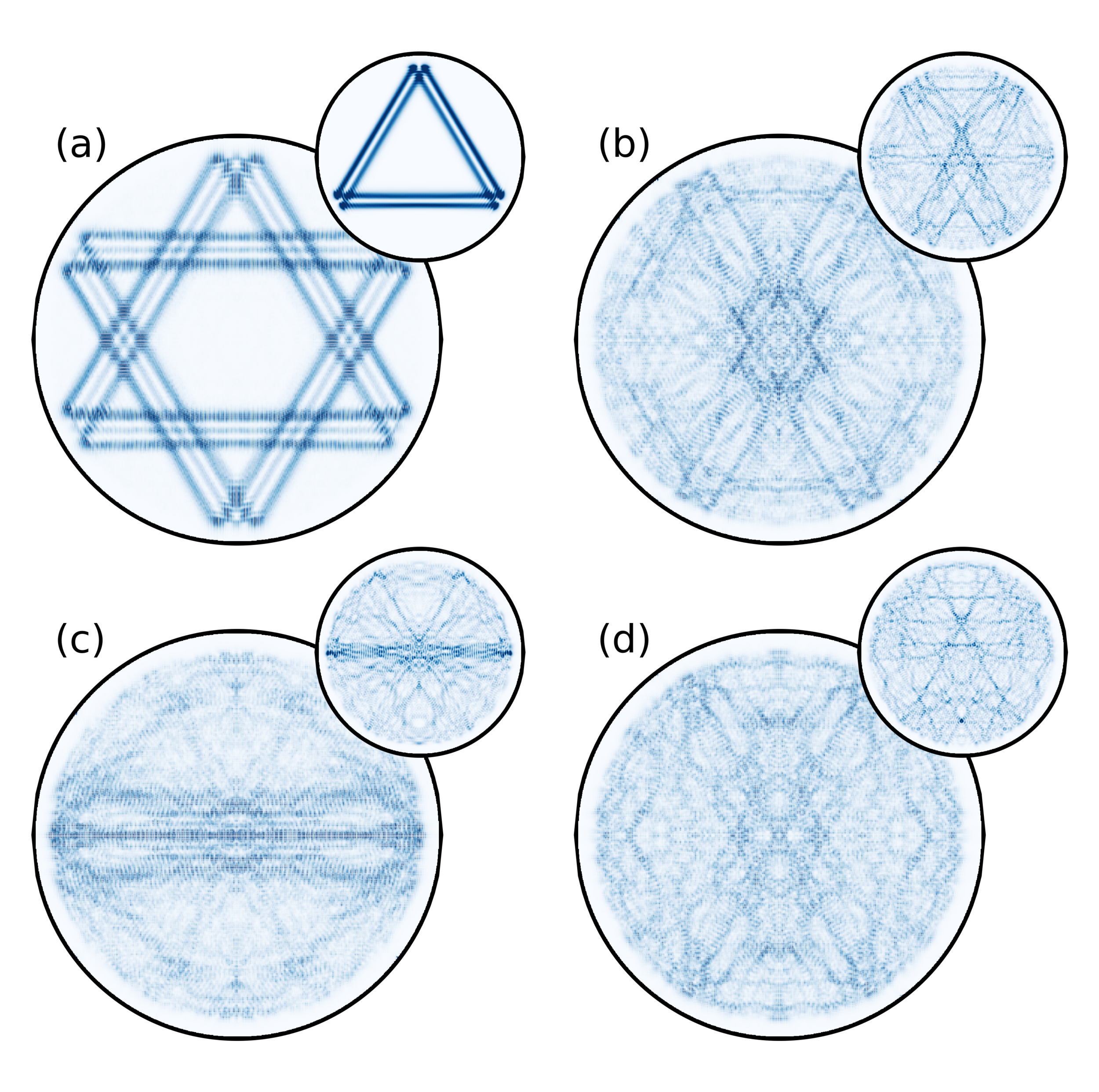}
    \caption{Probability density of selected eigenstates ($E_n = 100\pm 0.1$ meV) of a closed BLG cavity with $U = 40$ meV and $\mu_c = 100$ meV. We show examples for (a)-(b) regular quantum states, (c) a "bouncing ball" scar, and (d) an "ergodic state". Large plots show the LDOS when  both valleys are included due to fermion doubling, the small insets show related probability densities when only one valley contributes.}
    \label{fig:LDOS_DISCUSSION}
\end{figure}

To simulate the electronic states within the cavity, we use the tight-binding python module \textit{kwant} \cite{grothKwantSoftwarePackage2014}, for which we discretize \cite{chenFourbandEffectiveSquare2024} the Hamiltonian in \eqn\eqref{eqn:HBLG} on a square lattice with lattice spacing $b = 1$ nm.
This implementation reduces the size of the scattering matrix by a factor of $\sim 19$ compared to an atomistic tight-binding lattice and enables us to explore large system sizes with ballistic transport properties.
The discretization on the square lattice introduces numerical artifacts, namely fermion doubling and an additional deformation of the Fermi contour by the $D_4$ symmetry of the square lattice.
As for the fermion doubling, since the smooth electrostatic confinement potential of \eqn\eqref{eq:Transition Function} ensures inter valley scattering to be strongly suppressed, the additional valleys can be used to simulate both $K^\pm$ valleys simultaneously. To mitigate the influence of the artificial $D_4$ symmetry breaking, we restrict all calculations to energies below $150$ meV for which the deformation of the Fermi lines is negligible. We discuss details of our numerical implementation in   Appendices \ref{App:DiscLattice} to \ref{App: Rotating the Hamiltonian and Wilson Mass Term}.

We exemplify the probability densities of a closed cavity with an electrical displacement field $U=40$ meV and an energy band offset $\mu_c = 100$ meV for a selection of eigenstates in Fig. \ref{fig:LDOS_DISCUSSION}.
Here, large plots correspond to eigenstates in which both valleys contribute simultaneously. The smaller insets show eigenstates of the system where only the $K^-$ valley has been taken into account. We achieved the latter separation of valleys by adding a Wilson mass term \cite{staceyEliminatingLatticeFermion1982} to gap out the additional valleys (for details see Appendix \ref{App:FermionDoublingWillsonMass}).
Previous ray tracing calculations \cite{seemannGatetunableRegularChaotic2023} as well as our own semiclassical calculations (see below and Appendix \ref{App : Closed Cavity}) have shown that the closed cavity exhibits a mixed phase space, containing chaotic as well as regular regions.

\begin{figure}
    \centering
    \includegraphics[width=0.9\linewidth]{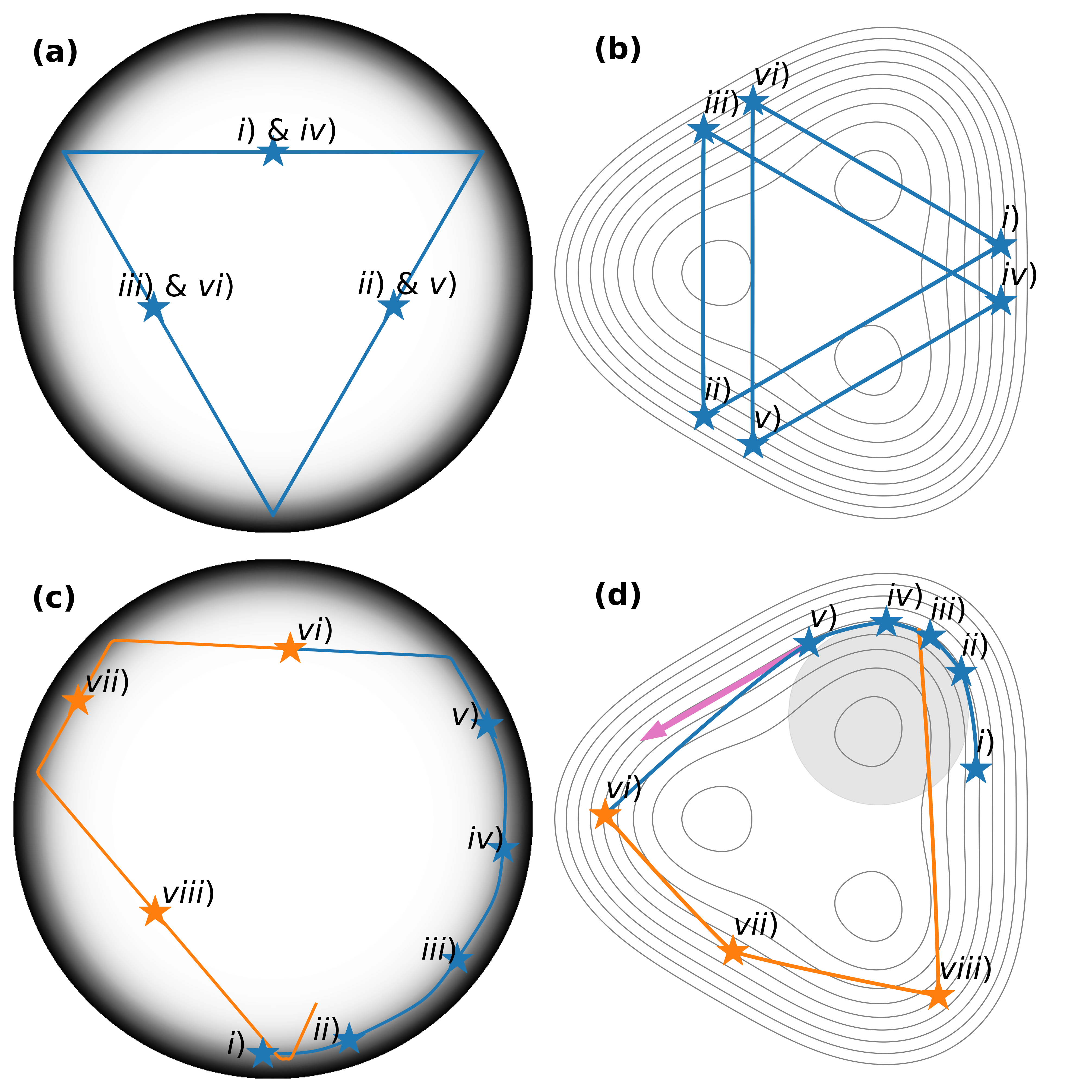}
    \caption{(a) Real space  semiclassical trajectory of a stable periodic orbit (blue) in a circular BLG cavity with smooth boundaries. Propagation through the cavity is numbered $i)-vi)$. (b) Fermi lines ($K^-$ valley) and  k-space trajectory of the stable periodic orbit in momentum space. The scattering events at the boundary are given by the (blue) lines connecting the stars, marking momenta propagation ($i)-vi)$). Straight lines in k-space imply the conservation of the momentum parallel to the local scattering area. (c) Trajectory that starts as a whispering gallery-like mode (blue) and  transitions to a standard trajectory (orange). Propagation ($i)-ix)$) corresponds to the star marked points in k-space (d). For the k-space trajectory connecting $i)-v)$ the combination of the circular cavity and the locally rotational invariant k-space region (gray circle), results in a local conservation of angular momentum.
    Scattering events occurring over larger areas within the smooth boundary lead to curved k-space trajectories.
    }
    \label{fig:scl_wisp}
\end{figure}

Accordingly, we classify the cavity eigenstates into three categories.
Firstly, we find regular quantum states corresponding to stable periodic orbits in the regular region. For these states, the probability density generally shows triangular patterns enhanced around the stable classical orbits exhibiting $D_6$ symmetry, cf.~\fig\ref{fig:LDOS_DISCUSSION} (a) and (b).
The morphology and energies of these eigenstates can be explained by the Einstein-Brillouin-Keller quantization rules \cite{ClosedOrbitsRegular1976} 
as they correspond to stable triangular orbits induced by the trigonal warped BLG dispersion \cite{seemannGatetunableRegularChaotic2023}.
We visualize a semiclassical electron trajectory of such a stable periodic orbit in \fig \ref{fig:scl_wisp} (a). We connect the real-space trajectory inside the cavity (blue line in \fig \ref{fig:scl_wisp}(a)) and its propagation in momentum space (\fig \ref{fig:scl_wisp}(b)). The free propagation between points $i)-vi)$ in real space corresponds to stationary points in momentum space marked by stars.
For this triangular trajectory, the scattering events correspond to straight lines, implying the conservation of the momentum parallel to the local scattering area.
Hence, scattering events occur in such a small real-space region that the curvature of the cavity is negligible and the trajectories are similar to those obtained for hard wall boundary conditions (compare \cite{seemannGatetunableRegularChaotic2023}).

The second class of eigenstates emerges from short unstable periodic orbits positioned in the chaotic region.
Quantum states enhanced along such orbits are commonly known as quantum scars \cite{hellerBoundStateEigenfunctionsClassically1984, keski-rahkonenVariationalScarringGraphene2025, geDirectVisualizationRelativistic2024}. In \fig\ref{fig:LDOS_DISCUSSION} (c), we exemplify a "bouncing ball" scar \cite{selinummiFormationPrevalenceStability2024, chalangariVariationalScarringOpen2025} corresponding to the straight back and forth motion of a classical particle within the cavity \cite{seemannGatetunableRegularChaotic2023}.

Finally, all other states we label ergodic, as they live 
in the chaotic region of the phase space but do not display scarring (cf.~the example in \fig\ref{fig:LDOS_DISCUSSION} (d)).

 The quantum solutions in a closed cavity obtained here from our tight-binding calculations confirm the predictions of the classical ray tracing simulations with respect to stable, triangular orbits, bouncing ball modes, and the ergodic regime \cite{seemannGatetunableRegularChaotic2023}.
 However, we were unable to identify whispering gallery-type modes \cite{zhaoCreatingProbingElectron2015, brunGrapheneWhisperitronicsTransducing2022,ramanWhisperingGalleryPhenomenaSt1921,zhengCoexistenceElectronWhisperinggallery2022} in our tight-binding solutions for systems with smooth electrostatic confinement described by \eqn\eqref{eq:Transition Function} (compare also \cite{emmanielenQuantumScarsBilayer2023}).

 To investigate the absence of whispering-gallery-type modes in our tight-binding calculations, we complement our analysis with simulations under hard-wall boundary conditions for the tight-binding model (see Appendix~\ref{App : Closed Cavity}) and smooth electrostatic confinement for the trajectory-based approach.
 For the latter we employed the semiclassical equations of motion \cite{neilw.ashcroftFestkorperphysik2001},
\begin{align}
    \mathbf{\dot{r}} = \frac{\delta H }{\delta \mathbf{p}}, \quad \mathbf{\dot{p}} = -\frac{\delta H}{\delta \mathbf{r}} ,\label{eq:SCL_EOM}
    \intertext{with the Hamilton function}
    H(\mathbf{r},\mathbf{p}) = \epsilon \left(\frac{\mathbf{p}}{\hbar} \right) - V(\mathbf{r}),
    \label{eq:SCL_EOM_2}
\end{align}
and the confining potential $V(\mathbf{r}) = \mu(\mathbf{r})+\mu_c$.
Our analysis reveals that whispering-gallery-type boundary modes are present among the cavity states unless the system features both a smooth cavity boundary and a trigonal distortion of the Fermi contour (for detailed comparison between different degrees 
of trigonal warping see Appendix~(\ref{App : Closed Cavity})).

To understand the disappearance of the whispering gallery states in such realistic BLG cavities including both a smooth confinement and the trigonally warped dispersion, we follow an exemplary semiclassical trajectory in real space (Fig.\ref{fig:scl_wisp} (c)) and in \textit{k}-space (Fig.\ref{fig:scl_wisp} (d)) simultaneously.
The trajectory begins in a whispering-gallery-like fashion (staying close to the cavity boundary for several scattering events, colored in blue) before abruptly transforming into a standard trajectory (orange).
For sharp incidence angles of the real-space trajectories to the cavity boundary, the connecting lines in momentum space are straight,
indicating a hard-wall-like scattering process where the momentum parallel to the hard-wall boundary is conserved.
For flat incident angles, the \textit{k}-space trajectory bends due to the smooth curved boundary, breaking the conservation of the parallel momentum component.
In the vicinity of the approximately circular corners of the triangular Fermi contours, the local (in the sense of local in phase space) angular momentum is conserved, rendering the whispering-gallery states locally stable.
Once the trajectory leaves the quasi-circular corners, hard wall boundary conditions allow skipping the flat sides of the triangularly warped Fermi contour (pink arrow in Fig. \ref{fig:scl_wisp} (b)). Re-entering another locally circular Fermi-contour region, corresponding to a shallow real-space incident angle, stabilizes the gallery mode in this case. 
In contrast, a smooth boundary bends the \textit{k}-space trajectory, resulting in a momentum- and real-space combination that no longer supports whispering-gallery modes.


In summary, all existing eigenstates (see Fig. \ref{fig:LDOS_DISCUSSION}), independent of their classification, display spatial structures of enhanced probability density that follow the dominant group velocities induced by the triangularly warped Fermi surface.
In the following, we will examine how these preferred directions influence the transmission properties of the BLG cavities.


\section{Anisotropic Transport through a Bilayer Graphene cavity\label{sec:AnisotrpicTransport}}
\begin{figure}
    \centering
    \includegraphics[width=\linewidth]{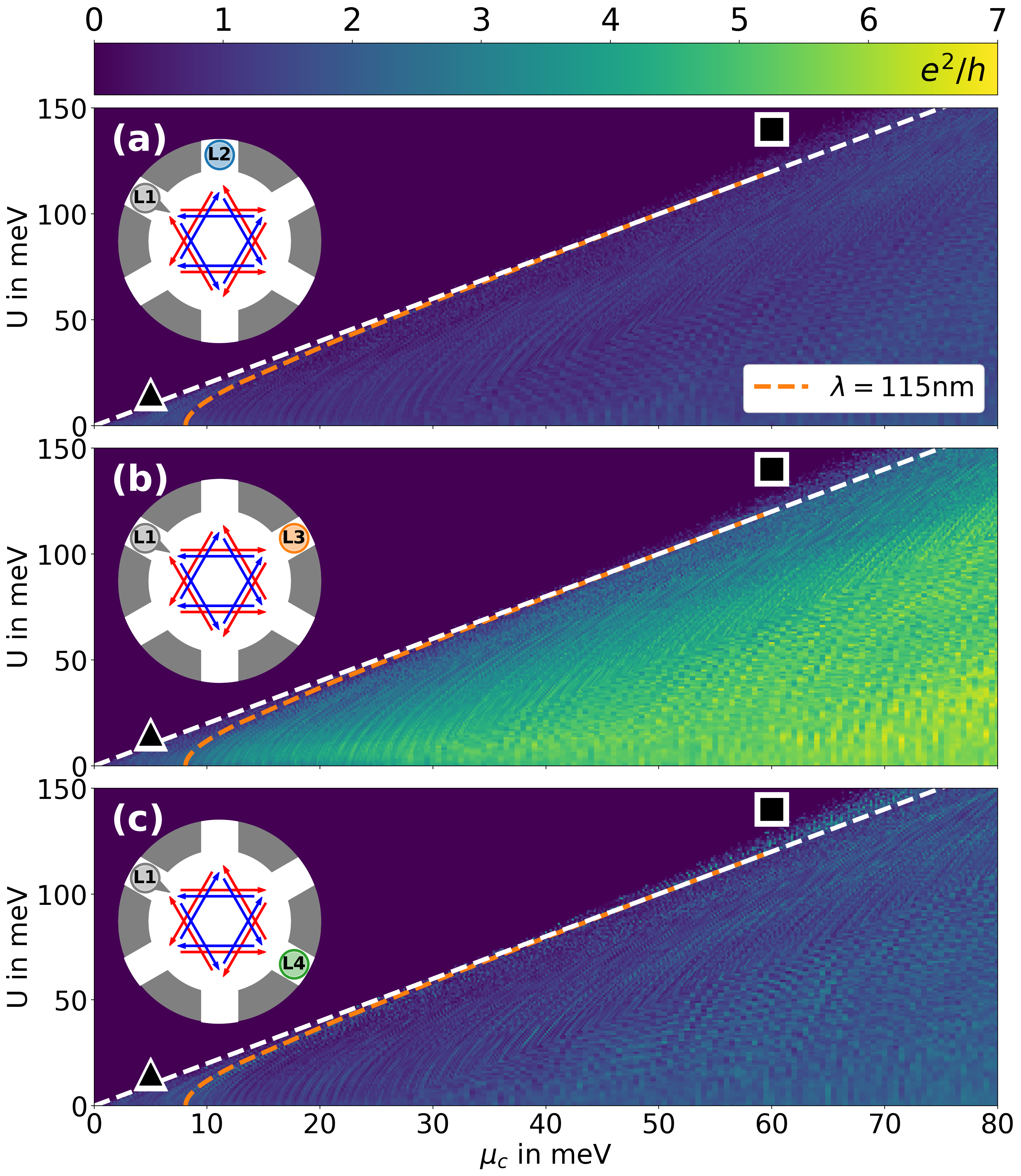}
    \caption{Conductance through the BLG cavity for (a) neighboring lead pairs, (b) triangular connected lead pairs and (c) opposing lead pairs depending on the displacement field $U$ and the band offset $\mu_c$. The orange dashed line marks the Fermi wavelength $\lambda_F = 115$ nm.     
    The white dashed line denotes the Lifshitz transition.}
    \label{fig:ConductanceHeatmap}
\end{figure}
To study transport through the BLG cavity, we attach leads to the device. We chose a setup with six leads of $50$nm width, each separated by $60^\circ$ degrees, as sketched in \fig \ref{fig:ConductanceSetupBandstructure}(a). This geometry reflects the $D_6$ symmetry of the system, as prescribed by the dominant group velocities. We demonstrate later in the manuscript that such device design is ideally suited for measuring anisotropic transport as it allows to detect anisotropy independently of the orientation of the device to the lattice, cf.~\sect\ref{sec:AnisotrpicTransport}. 
To ensure multiple modes coupling to the system, the leads' band offset is fixed to $\mu_L = 200$ meV and transitions smoothly to the potential landscape of the cavity (for details see Appendix~\ref{App : Transport Calculations})
\footnote{As our focus lies on the transport properties of the cavity, the minor deformations due to the underlying square lattice of the Fermi lines inside the leads can be dismissed.}.
    
Here, we focus on one specific device-to-lattice orientation, $\alpha = 0^\circ$ in \eqn\eqref{eqn:pi}, for which we sketch the six propagation directions in \fig\ref{fig:ConductanceSetupBandstructure}.
We calculate the conductance through the cavity for a variety of displacement strengths $U \in [0\text{meV},150\text{meV}]$ and band offsets $\mu_c \in [0\text{meV},100\text{meV}]$.
We present the conductance data for the lead pairs $(L1,L2)$ (neighboring), $(L1,L3)$ (triangular connected), $(L1,L4)$ (opposing) in \fig \ref{fig:ConductanceHeatmap}. 
Note that all other lead combinations are included in this selection as the symmetry of the system ensures that all neighboring lead pairs are equivalent to the lead pair $(L1,L2)$, all lead pairs connected by the group velocities can be matched to the lead pair $(L1,L3)$ and all opposing lead pairs are represented by the pair $(L1,L4)$.

The most prominent feature of \fig\ref{fig:ConductanceHeatmap} is the strongly enhanced conductance between triangular connected leads $(L1,L3)$ compared to the other lead combinations.
The strong conductance between these leads can be traced back to the triangular warping of the Fermi contour: leads connected via the dominant group velocities exhibit high conductance, while the conductance of all other lead combinations is suppressed (cf.~also \fig\ref{fig:ConductanceSetupBandstructure}(c)).
We observe such a substantial anisotropic conductance enhancement  for almost all tested combinations of band gaps $U$ and band offsets $\mu_c$.
In Appendix~\ref{App : Transport Calculations}, we also demonstrate the robustness of the conductance enhancement against local scatterers in various conductance tests in the presence of white noise (see Fig. \ref{fig:DissorderSimualtions}).

Conversely,  conductance between triangularly connected leads is not enhanced in the two regions marked by $\blacktriangle$ and $\blacksquare$ in Fig. \ref{fig:ConductanceHeatmap}.
\begin{figure}
    \centering
    \includegraphics[width=\linewidth]{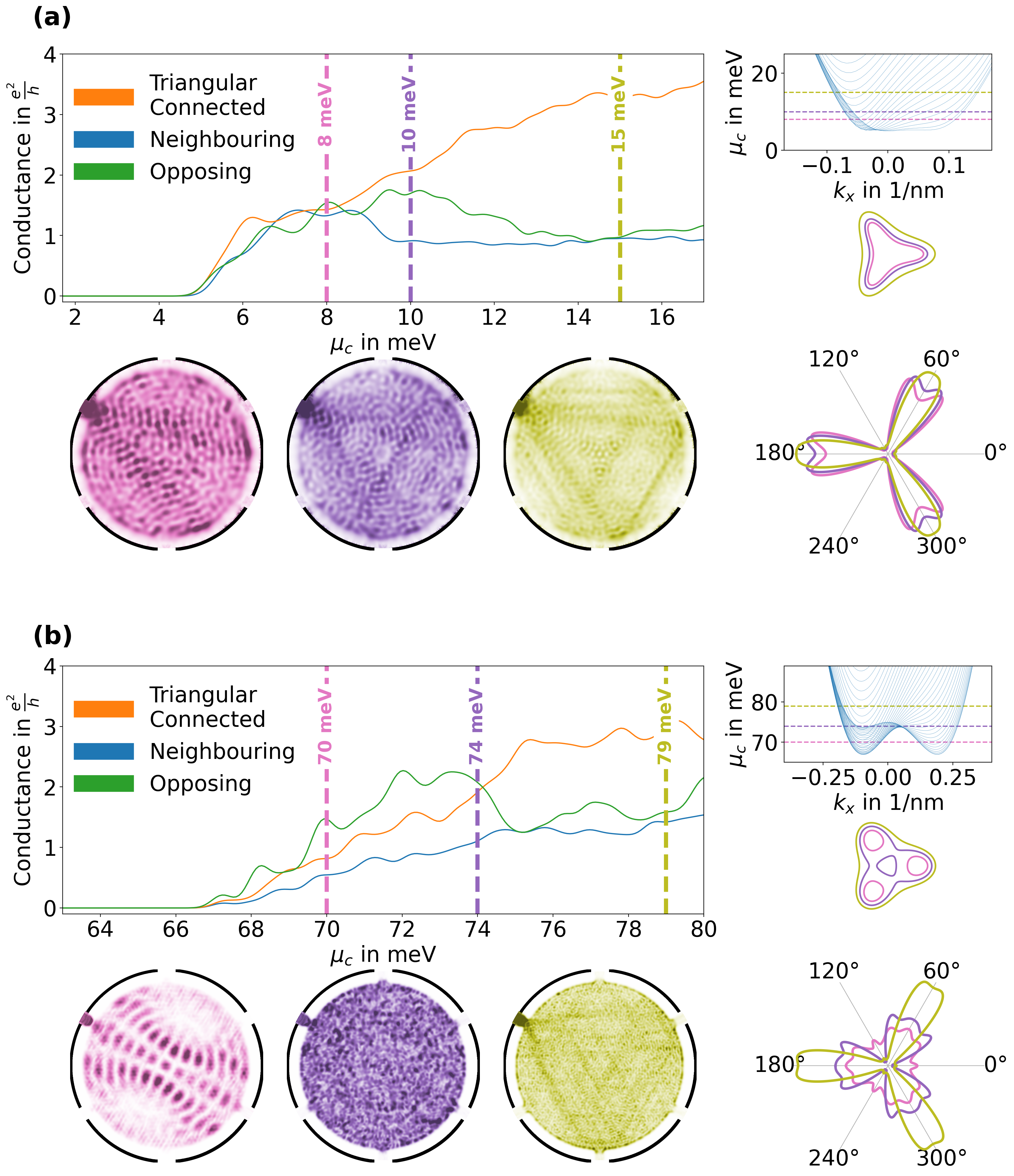}
    \caption{Thermal broadened - 2 K - line cuts at (a) $U=10$ meV and (c) $U=150$ meV through the conductance map of Fig.\ref{fig:ConductanceHeatmap}.
    (b),(d) mode-averaged current densities  for selected band offsets $\mu_c$.
    At the top of the outer right column the 2D-projections of the band-structure near the $K^+$ valley for the two different displacements fields, $U$, are displayed. The horizontal lines correspond to the dashed lines in the conductance plots. 
    Below the band-structures we show the Fermi-lines and corresponding group velocities.}
    \label{fig:LifshitzTransitionCloseUp}
\end{figure}

In the $\blacktriangle$-regime of small band offsets $\mu_c$ and displacement fields $U$, the Fermi wavelength $\lambda_F$ is too large for the modeled device to be considered ballistic, resulting in wave-governed transport.
The wave-governed transport entails conductance values of similar magnitude for all lead combinations despite anisotropic velocity distributions, visualized in the conductance cuts at $U=10$meV in Fig. \ref{fig:LifshitzTransitionCloseUp}(a).
With an increasing band offset $\mu_c$, the Fermi wavelength decreases until ballistic transport influenced by the triangular Fermi contour takes over. 
This transition, at which ballistic transport becomes dominant, manifests as a conductance drop-off between neighboring leads $(L1, L2)$. We mark this transition by an orange dashed line at $\lambda_F \approx 115$ nm in \fig \ref{fig:ConductanceHeatmap}. From this characteristic wavelength, it can be estimated that the system must be approximately seven times the Fermi wavelength to observe anisotropic ballistic transport reflecting the dominant group velocities.


Conversely, in the region marked by  $\blacksquare$ in \fig \ref{fig:ConductanceHeatmap}, the conductance between the opposing leads $(L1,L4)$ is larger than for any other lead combination. Here, since this enhancement is only seen for opposing leads, it cannot be explained by purely wave-governed transport. Instead,  transport characteristics in this regime reflect the Lifshitz transition of the BLG band structure, where the single, triangular Fermi lines break into individual pockets around the minivalleys \cite{varletAnomalousSequenceQuantum2014,knotheInfluenceMinivalleysBerry2018} (marked by the white dashed line in \fig \ref{fig:ConductanceHeatmap}).
We demonstrate the influence of the Lifshitz transition on transport, velocity distributions, and current densities in \fig\ref{fig:LifshitzTransitionCloseUp}(b) for a large displacement field of $U=150$ meV.
Here, the conductance between opposing leads (green) generally exceeds the conductance of the triangular connected leads (orange) for band offsets below the Lifshitz transition. In this regime, the Fermi contour is given by three separated mini-valleys, \fig \ref{fig:LifshitzTransitionCloseUp}, whose group velocity distribution shows less pronounced preferred directions compared to the singly connected triangular Fermi line. The corresponding mode-averaged current density connects the opposing leads, enabling direct transport through the cavity.
For band offsets above the Lifshitz transition, the six dominant group velocities form, leading to anisotropic transport properties through the cavity. Note that tuning between the distinct shapes of the BLG Fermi lines via the Lifshitz transition also changes the cavity dynamics from regular (below the Lifshitz transition) to chaotic (above the Lifshitz transition). We elaborate this point in detail in Appendix \ref{subsec:TuningIC}.




    \subsection{Magnetotransport}
    \begin{figure}
        \centering
        \includegraphics[width=1.\linewidth]{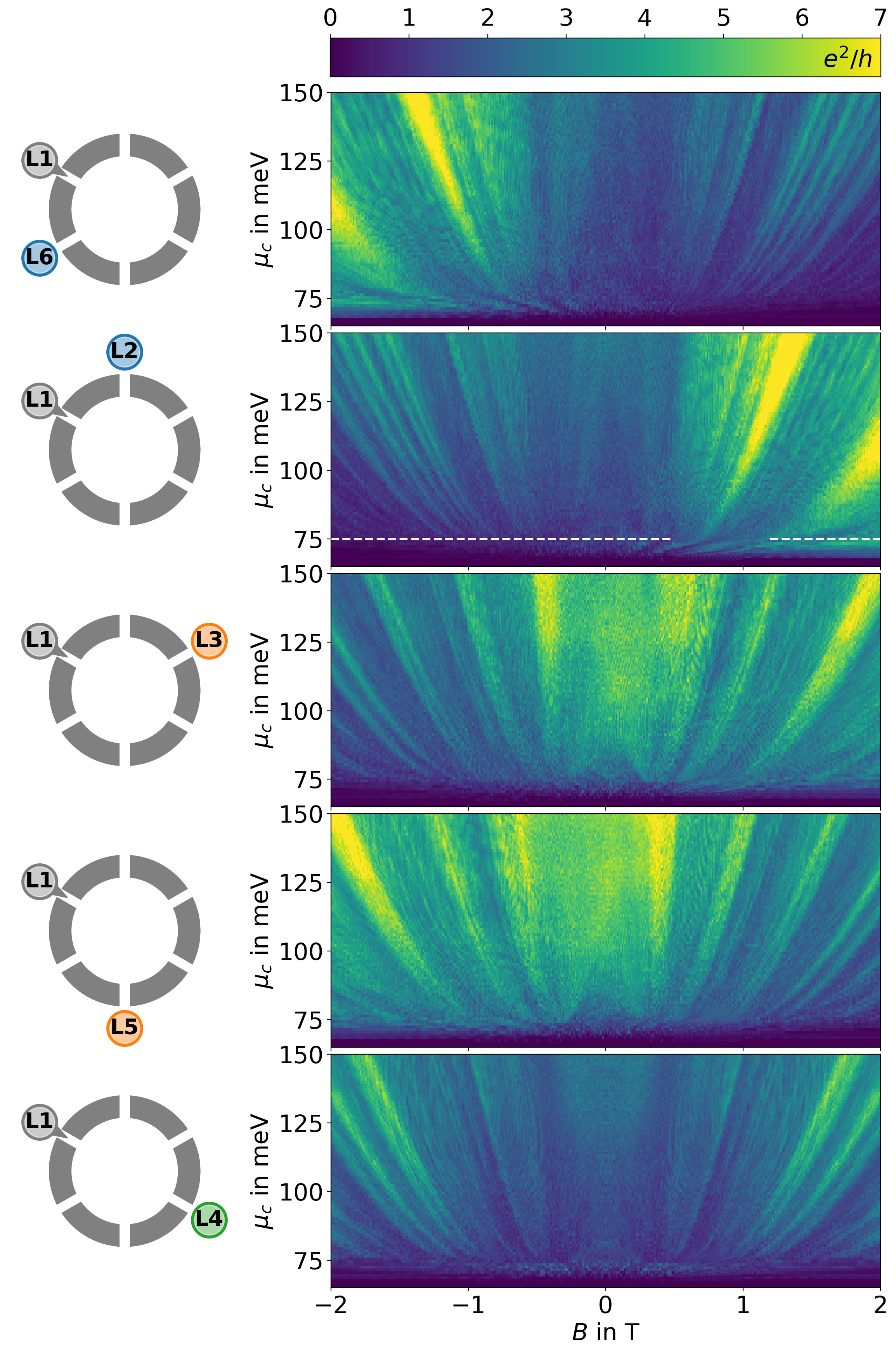}
        \caption{Conductance through the cavity for different lead combinations depending on the magnetic field and the potential landscape of the cavity. The regions of high intensity at larg magnetic fields arise from cyclotron motion. The kink in the high-intensity lines between $70$ meV and $80$ meV corresponds to the Lifshitz transition (marked by the white dashed line), reflecting the abrupt change in the cyclotron orbits' size. For the triangularly connected lead pairs, a conductance plateau emerges at higher Fermi energies when the flat segments of the triangularly warped cyclotron orbits become sufficiently large to traverse the cavity before bending.}
        \label{fig:Magnetotransport}
    \end{figure}
We now investigate the influence of a perpendicular magnetic field $\mathbf{B} = (0,0,-B)^T$ on the anisotropic transport through the BLG cavity.     In Fig. \ref{fig:Magnetotransport}, we present conductance data for magnetic fields from $-2 $ T to $2$ T and band offsets $\mu_c$ between $65$ meV and $150$ meV, assuming an incoming wave from lead $L1$. To capture effects due to the Lifshitz transition, we fix the displacement field at $U=150$ meV.
    
For all lead combinations, we observe finite conductance above the band gap and unevenly curved, high-intensity lines for sufficiently large magnetic fields. 
The latter can be associated with skipping orbits, focusing the electrons into leads for specific combinations of band offset and magnetic field. Here, the onset of the Lifshitz transition (marked by the white dashed line) can be identified by a kink in the slope of these lines, due to the abrupt change of electronic orbits' size. 
    
Most strikingly, we observe a plateau of enhanced conductance for finite magnetic fields $-0.5\text{T} \lesssim B \lesssim 0.5 \text {T}$ for the triangular connected leads $(L1,L3)$ and $(L1,L5)$ at sufficiently high Fermi energies above the Lifshitz transition. In this regime, the flat sides of the triangular cyclotron orbits still connect the leads over a finite range of $B$ before they begin to curve under the influence of the magnetic field. Hence, the triangular shape of the Fermi surface stabilizes the anisotropic transport up to a critical magnetic field strength \cite{kraftAnomalousCyclotronMotion2020}. We verify in Appendix \ref{App : Magneto Transport} that such a plateau is indeed not present in the case of a circularly symmetric dispersion with round cyclotron orbits. 
    

    \subsection{Rotating of the underlying lattice} \label{Sec: Rotating the lattice}
    In the previous conductance calculations, we assumed a specific orientation of the device with respect to the BLG lattice that allows for a direct path between triangularly connected leads when following one of the six dominant group velocities. 
    However, the orientation of the lattice is generally unknown in any real experiment. Therefore, we now demonstrate that the $D_6$ symmetry of the six-lead setup we chose ensures anisotropic transport measurements independent of the lattice-to-device orientation.
    To simulate a continuous rotation of the BLG lattice, we rotate the continuum model four-band-Hamiltonian clockwise using a finite rotation angle $\alpha$ in \eqn\eqref{eqn:pi}, before discretizing it on the square lattice.
    In contrast to the previous calculations with $\alpha =0^\circ$, for finite $\alpha$ a  Wilson mass term \cite{staceyEliminatingLatticeFermion1982} must be used to remove the additional valleys introduced by fermion doubling, as the nonphysical valleys rotate in the opposite (counterclockwise) direction (compare \fig \ref{fig:WillsonMassInfluence} in Appendix \ref{App: Rotating the Hamiltonian and Wilson Mass Term}). 

\begin{figure}
            \centering
            \includegraphics[width=\linewidth]{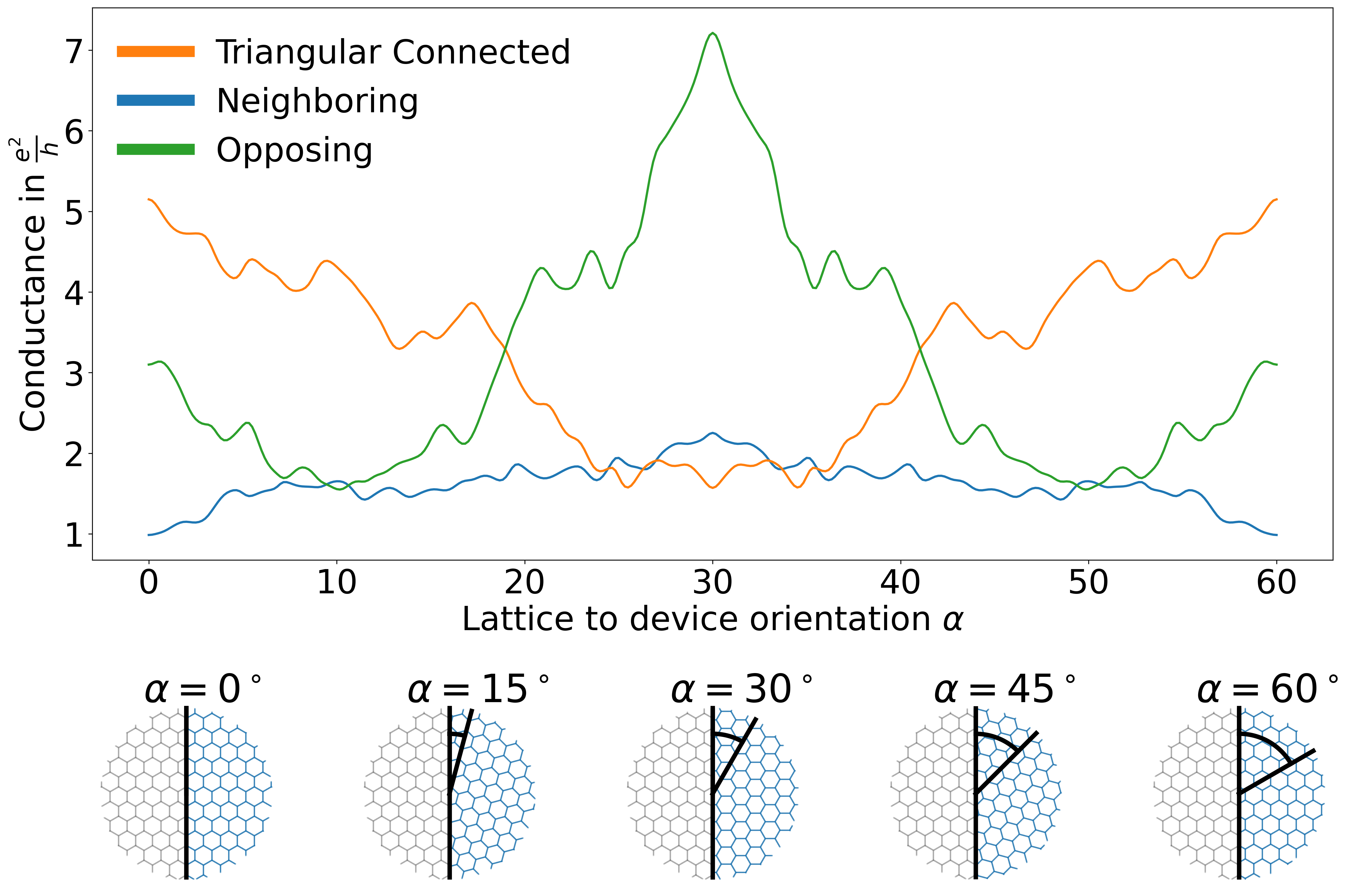}
            \caption{Averaged conductance for all neighboring (blue), triangular connected (orange) and opposing (green) lead combinations for $U=40$ meV and $\mu_c = 60$ meV depending on the lattice to device orientation given by the rotation angle $\alpha$ (see Eq. \ref{eqn:pi}). The definition of the rotation angle is visualized below, where the standard orientation used for all other calculations is denoted by $\alpha = 0^\circ$.}
            \label{fig:RotatingConductance}
        \end{figure}

    We show the conductance through the cavity for both valleys at $U=40$ meV and $\mu_c=60$ meV  in Fig. \ref{fig:RotatingConductance}.
    While the conductance between different lead pairs varies with the rotation angle $\alpha$ between the lattice and the device, the three different classes of lead combinations (triangular connected, neighboring, and opposing) exhibit clearly distinguishable conductance characteristics for all lattice orientations. 
    The conductance peak for opposing leads at $\alpha = 30^\circ$ follows from the ballistic nature of the transport, as in this case the dominant group velocities connect the opposing leads by traveling straight through the cavity. 
    As the system as well as the lattice properties are invariant under a $60^\circ$ degree rotation, the transmission is $\pi/3$-periodic in $\alpha$.
    We hence conclude that a setup featuring six leads, as proposed in this work, will exhibit strong anisotropic transport characteristics regardless of its orientation relative to the BLG lattice.
    The ratios between the different measured conductance values may provide evidence of the lattice-to-device orientation.

    The stability of the simulations with respect to noise, the robustness of the anisotropic measurements against arbitrary lattice orientations,
    and the simple geometry of the cavity, along with experimentally feasible system sizes, makes this proposed device design a promising platform for experimental studies.


\section{Conclusion}
Overall, our results demonstrate that the properties of gate-defined electronic BLG cavities reflect the material's rich, anisotropic electronic structure. We identify triangular modes in the quantum solutions of a closed, circular cavity induced by the triangularly warped Fermi surfaces. We verify the stability of such triangular modes with respect to system openings when attaching leads, entailing anisotropic transport through the cavity. The triangular Fermi contours also stabilise said anisotropy in the presence of small magnetic fields. When investigating the energy dependence of the transconductance, we further observe signatures of BLG's Lifshitz transition in the transport characteristics. Finally, we present an experimentally feasible setup that allows for the detection of the transport anisotropy independently of the orientation of the device with respect to the BLG lattice and to obtain controllable, preferred directions of transport.

We hence propose electronic transport as a versatile tool to study two-dimensional materials-based electronic cavities alternative to microscopy \cite{keski-rahkonenVariationalScarringGraphene2025,geDirectVisualizationRelativistic2024,cabosartRecurrentQuantumScars2017}. Concurrently, circular, gate-defined cavities allow for the generation of direction-modulated transport.
Unlike monolayer graphene \cite{zhangTransportSignaturesRelativistic2020,huangRelativisticQuantumScars2009,xuChaosDiracElectron2018,cabosartRecurrentQuantumScars2017}, BLG allows for electrostatic confinement by the opening of a bandgap.
Compared to scar-mediated anisotropic transport proposed in earlier works \cite{selinummiFormationPrevalenceStability2024,chalangariVariationalScarringOpen2025,luukkoStrongQuantumScarring2016,keski-rahkonenControllableQuantumScars2017,zijderveldQuantumScarsCaustics2024,keski-rahkonenVariationalScarringGraphene2025,zozoulenkoQuantumScatteringResonant1997,kimWaveFunctionScarring2002,birdLeadOrientationDependentWaveFunction1999,wisniackiScarringOpenQuantum2008}, where one manipulates the directionality using local
impurities, perturbations induced by tip potentials, or the cavity shape, in BLG, the anisotropy stems from the trigonally warped band structure itself. Electron cavity dynamics in anisotropic materials is hence particularly rich, offering opportunities to observe exotic bound states and to utilize them for direction-dependent transport.

\section{ACKNOWLEDGMENTS}
We would like to thank Lukas Seemann and Martina Hentschel for discussions and their valuable insights. We thank Michael Barth and Michael Wimmer for guidance at an early stage of the work. Further, we acknowledge enriching discussions about experimental feasability with Christoph Stampfer, Christian Volk, Hubert Dulisch,  Simone Sotgiu, and Carolin Gold. We acknowledge funding by the Deutsche Forschungsgemeinschaft (DFG, German Research Foundation) within  DFG Individual grant KN 1383/4 (Project-ID 529637137) and SFB 1277, Project A07 (Project-ID 314695032).

\clearpage

\appendix

\section{Discretisation of the continous Hamiltonian} \label{App:DiscLattice}
To discretize the Hamiltonian in Eq. \ref{eqn:HBLG} onto a square lattice 
\begin{align}
    \mathbf{R}_{n_1,n_2} = n_1 (b\hat{e}_x) + n_2 (b\hat{e}_y) \quad ;n_1,n_2 \in \mathbb{Z},
\end{align}
with the lattice constant $b$ and the unit vectors $\hat{e}_{x/y}$ in x and y direction, one applies the central finite difference
\begin{align}
    &p_x \rightarrow -\frac{i\hbar}{2b}(\Psi_{n_1+1,n_2}-\Psi_{n_1-1,n_2})  \label{eq: px_disc}\\
    &p_y \rightarrow -\frac{i\hbar}{2b}(\Psi_{n_1,n_2+1}-\Psi_{n_1,n_2-1}),
\end{align}
where $\Psi_{n_1,n_2}$ denotes the wave function at the lattice point $\mathbf{R}_{n_1,n_2}$.
This results in the following matrices for the tight-binding hopping and onsite terms:
\begin{align*}
    H_\text{onsite} &= \left( \begin{array}{cccc}
        \epsilon_{A1} & 0 & 0 & 0 \\
        0 & \epsilon_{B1} & \gamma_1 & 0 \\
        0 & \gamma_1 &  \epsilon_{A2} & 0\\
        0 & 0 & 0 & \epsilon_{B2}
    \end{array} \right) \\ \\
    H_\text{x-hopping} &= -\frac{i\hbar}{2b} \left( \begin{array}{cccc}
         0 & v\theta^\dagger & -v_4 \theta^\dagger & v_3 \theta\\
         v \theta & 0 & 0 & -v_4 \theta^\dagger\\ 
         -v_4 \theta & 0 & 0 & v \theta^\dagger\\
         v_3 \theta^\dagger & -v_4 \theta & v \theta  & 0 
         \end{array} \right) \\ \\
    H_\text{y-hopping} &= -\frac{i\hbar}{2b} \left( \begin{array}{cccc}
     0 & v\phi^\dagger & -v_4 \phi^\dagger & v_3 \phi\\
     v \phi & 0 & 0 & -v_4 \phi^\dagger\\ 
     -v_4 \phi & 0 & 0 & v \phi^\dagger\\
     v_3 \phi^\dagger & -v_4 \phi & v \phi  & 0 
     \end{array} \right),
\end{align*}
where we introduced 
\begin{align*}
    \theta = \xi \cos(\alpha)+i \sin(\alpha)& \\
    \theta^\dagger = \xi \cos(\alpha)-i \sin(\alpha)& \\
    \phi = -\xi \sin(\alpha)+i \cos(\alpha)& \\
    \phi^\dagger = -\xi \sin(\alpha)- i \cos(\alpha)&,
\end{align*}
for short hand notations.

\section{Fermion Doubling and Deformation of Fermi Surfaces}\label{App:FermionDoublingWillsonMass}
\begin{figure}
    \centering
    \includegraphics[width=1\linewidth]{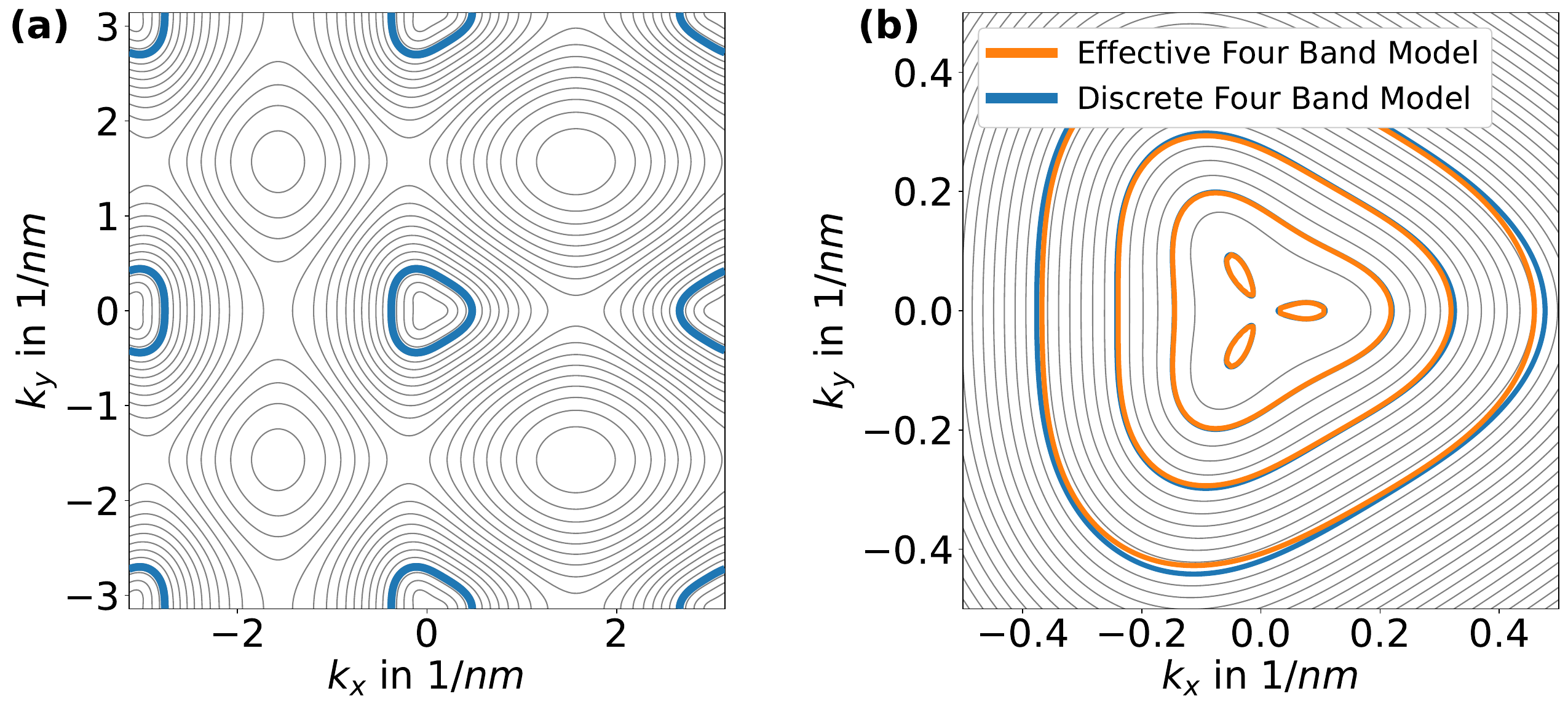}
    \caption{(a) Fermi contours of the discretized Hamiltonian for $U = 40$ meV, showing both
    the physical valley (center) and additional valleys due to fermion doubling, highlighted in blue at the Fermi energy $E_F = 150$ meV.
    The deformation caused by the lattice symmetry is clearly visible.
    (b) Close-up of the center valley, comparing Fermi contours from the continuous Hamiltonian (orange) and the discretized model (blue) at energies $20,40,80,150$ meV (inner to outer lines).
    The deviation between the models is negligible for energies up to $150$ meV, which holds true for all simulated values of $U$.}
    \label{fig:FourBandFermiLines}
\end{figure}

Discretizing the continuous model onto the square lattice causes fermion doubling at the edges and corners of the Brillouin-zone and imposes a $D_4$ symmetry on the band-structure (see \fig \ref{fig:FourBandFermiLines} (a)).
For non-rotated ($\alpha = 0^\circ)$ Hamiltonians, these additional valleys allow simulations of both BLG valleys simultaneously.
The lattice-induced deformation becomes more prominent for higher energies, limiting the accuracy of calculations.
Here we chose to stay bellow $150$ meV, as the deformation of the Fermi contour and its impact on the physical properties governing transport is small (see \fig \ref{fig:FourBandFermiLines} (b)). 

\section{Rotating the Hamiltonian and Wilson mass term} \label{App: Rotating the Hamiltonian and Wilson Mass Term}
\begin{figure}
    \centering
    \includegraphics[width=1\linewidth]{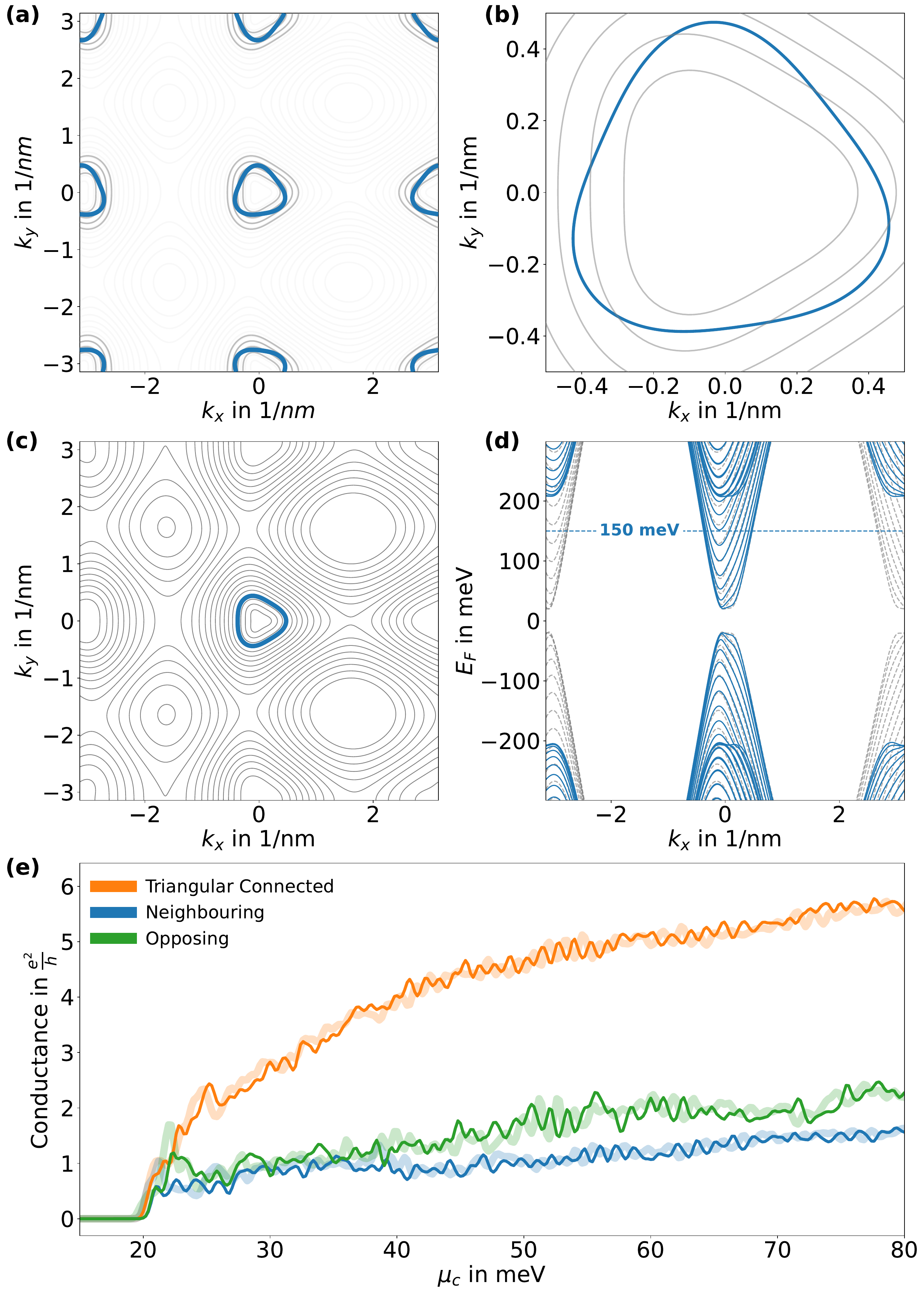}
    \caption{(a) In gray the low-energy regime of the Fermi surface of the non rotated Hamiltonian at $U=40$ meV . In blue, the Fermi surfaces at $E_F = 150$ meV for a rotated ($\alpha \neq 0^\circ$) Hamiltonian are displayed.
    When the central valley as well as the corner valleys are rotated clockwise, the valleys at the edges rotate counterclockwise. 
    (b) Close up of (a) focusing on the central valley.
    (c) Fermi contour for a system with a Wilson mass term of $w_D = 47$ meV.
    In blue, the Fermi contour for $E_F = 150$ meV. All non-physical valleys are removed.
    (d) Band structure of a BLG lead with $50$ nm width. In gray the band-structure without and in blue the band-structure including the Wilson mass term. 
    (e) Conductance through the circular cavity (compare \fig \ref{fig:ConductanceSetupBandstructure}) with a Wilson mass term with  $w_D = 47$ meV (thin solid lines) and without Wilson mass term (thick bright lines). For a Wilson mass parameter greater than zero, a separate calculation is performed for each valley, and the results are summed and multiplied by two to account for spin degeneracy.}
    \label{fig:WillsonMassInfluence}
\end{figure}

While the nonphysical valleys can be used to calculate two valley systems for non-rotated Hamiltonians, this approach breaks down once the Hamiltonian is rotated, e.g. clockwise ($\alpha > 0^\circ$), as the nonphysical valleys at the edges rotate counterclockwise (compare \fig \ref{fig:WillsonMassInfluence}(a) and \fig \ref{fig:WillsonMassInfluence}(b)).
One approach to avoid the nonphysical valleys is to use a Wilson mass term \cite{staceyEliminatingLatticeFermion1982} to increase the energies at the edges of the Brillouin-zone such that they lie far above the studied energy regime and do not contribute to the transport calculations.
In the case of the previously introduced tight binding setup, the additional Wilson mass can be included by adding
\begin{align}
    H_W = w_D \left( \begin{array}{cccc}
    1 & 0 & 0 & 0\\
    0 & -1 & 0 & 0\\
    0 & 0 & 1 & 0\\
    0 & 0 & 0 & -1
    \end{array} \right),
\end{align}
to the on-site and hopping matrices in the following manner
\begin{align}
    &H'_\text{onsite} = H_\text{onsite} - 4H_W \\
    &H'_\text{x-hopping} = H_\text{onsite} + H_W \\
    &H'_\text{y-hopping} = H_\text{onsite} + H_W.
\end{align}
The energy parameter $w_D$ governs the strength of the Wilson mass term.
The influence of the Wilson mass term on the Fermi lines and band structure for a strength of $w_D = 47$ meV can be seen 
in \fig \ref{fig:WillsonMassInfluence}(c) and \fig \ref{fig:WillsonMassInfluence}(d).
An exemplary conductance comparison between systems with and without with Wilson mass term for a rotation angle of $\alpha = 0^\circ$ is displayed in \fig \ref{fig:WillsonMassInfluence} (e).

\section{Transport Calculations} \label{App : Transport Calculations}
For all transport calculations we considered lead widths of $50$ nm and a local energy band offset inside the leads of $200$ meV.
This combination results in 9 propagating modes per valley for each spin species, giving 18 modes per valley and 36 incoming modes in total.

\subsection{Transport without trigonal warped Fermi surface}
\begin{figure}
    \centering
    \includegraphics[width=\linewidth]{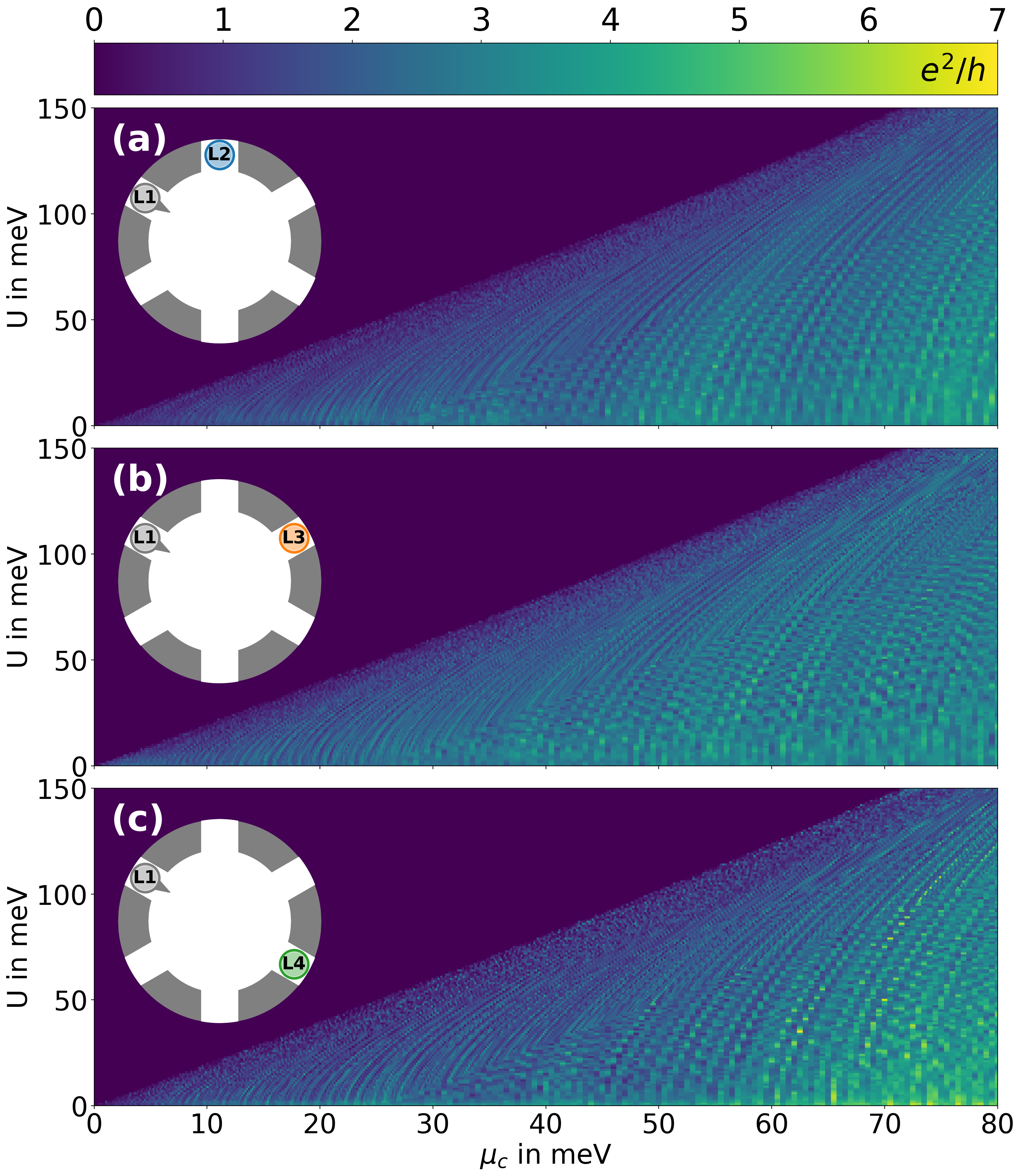}
    \caption{Same as \fig \ref{fig:ConductanceHeatmap} but without the triangular warped Fermi surfaces ($\gamma_3 = 0$ meV).}
    \label{fig:CompareHeatmapConductance}
\end{figure}

To investigate the effects of the trigonal warping, we also simulated the conductance through the BLG cavitiy for $\gamma_3 = 0$ meV, in which case the Fermi surfaces are circular.
As shown in \fig \ref{fig:CompareHeatmapConductance}, the anisotropic conductance features vanish. 

\subsection{Transport and local scatterers} \label{App : Noise and Transport}
\begin{figure}
    \centering
    \includegraphics[width=\linewidth]{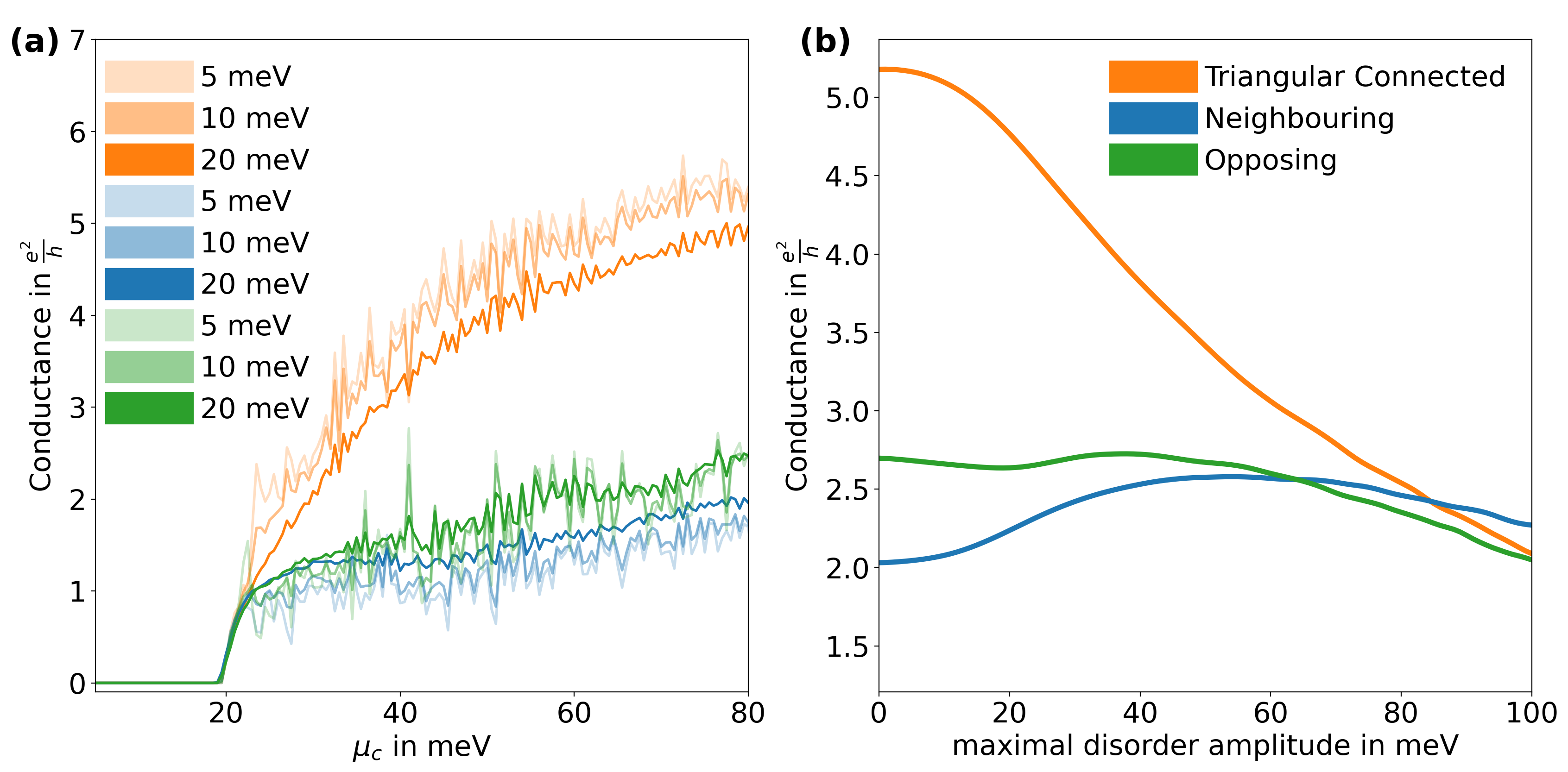}
    \caption{(a) conductance for different disorder strength (5 meV, 10 meV, 20 meV) at $U=40$ meV. Each curve consists of an average conductance over 100 different disorder calculations. (b) Plotted is the conductance at $U=40$ meV and $\mu_c = 80$ meV for increasing disorder amplitudes.}
    \label{fig:DissorderSimualtions}
\end{figure}

In \fig \ref{fig:DissorderSimualtions} we study the system's sensitivity to white noise potentials of different magnitudes.
\fig \ref{fig:DissorderSimualtions}(a) shows various conductance calculations at $U=40$ meV for different 
maximal noise amplitudes ($5,10,$ and $20$ meV).
For each amplitude, we calculate 100 different realizations and display their average conductance.
While fluctuations in the conductance data decrease and the overall conductance values are slightly damped, the anisotropic transport features remain clearly visible.

In \fig \ref{fig:DissorderSimualtions}(b), we fix the displacement field $U=40$ meV and the energy band offset $\mu_c$ to $80$ meV to identify the critical disorder amplitude needed to suppress anisotropic transport.
To do so, we increase the disorder strength constantly and simulate 20 different realizations for each strength. 
We observe the anisotropic transport to be robust against large white noise disorder up to $\frac{\mu_c}{2}$.

\subsection{Magnetotransport} \label{App : Magneto Transport}
\begin{figure}
    \centering
    \includegraphics[width=\linewidth]{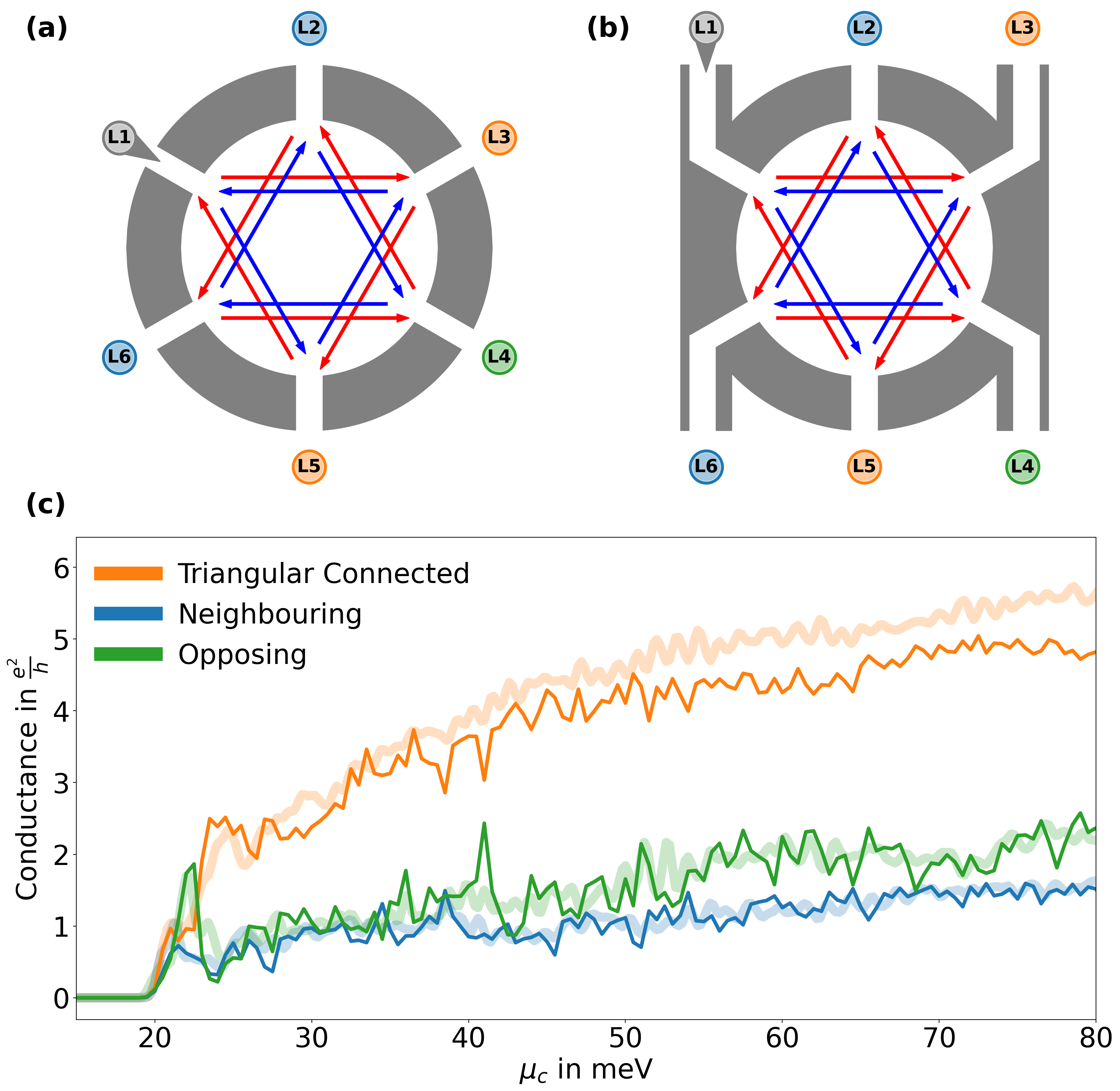}
    \caption{(a) Setup without magnetic field as shown in \fig \ref{fig:ConductanceSetupBandstructure}. (b) Schematic setup with additional lead connectors for the magneto-transport simulations. (c) Comparing the conductance between the setups displayed in (a) (bright thick line) and (b) (thin line) for $U=40$ meV and $B=0$ T.}
    \label{fig:CompareMagnetoTransport}
\end{figure}

To ensure translational invariance in the leads, the magnetic vector potential must align with the invariant direction of the semi-infinite leads. 
If multiple leads with different orientations are present, one may either smoothly vary the vector potential inside the cavity such that locally at each lead the translation invariance is satisfied \cite{mrenca-kolasinskaProbingMinibandStructure2023,barangerElectricalLinearresponseTheory1989}, or modify the scattering region, such that all leads share the same orientation (up to an optional sign change).
In this case we chose the latter approach as depicted in \fig \ref{fig:CompareMagnetoTransport}(b). 
As the new geometry adds additional perturbation, we verify that the anisotropic conductance behavior is still observable (compare \fig \ref{fig:CompareMagnetoTransport}).

To compare the influence of the triangular Fermi surfaces, we calculated the conductance for the magneto-transport for $\gamma_3 = 0$ meV, namely circular Fermi surfaces.
For high magnetic fields and large band offsets $\mu_c$ the structure of the plots in \fig \ref{fig:CompareMagnetoTransport} and \fig \ref{fig:Magnetotransport} are nearly identical.
For this parameter regime, the transport properties are dominated by small skipping orbits, and the exact form  of the skipping orbits -circular or trigonally warped- plays a minor role. 
For smaller magnetic fields, the conductance plateaus as well as the two regions of high conductance near the plateaus, a result of an auto-focusing effect- do not form in the absence of trigonal warping.
And finally, the prominent kink marking the Lifshitz transition is not visible in the case of $\gamma_3 = 0$ meV.

\begin{figure}
    \centering
    \includegraphics[width=\linewidth]{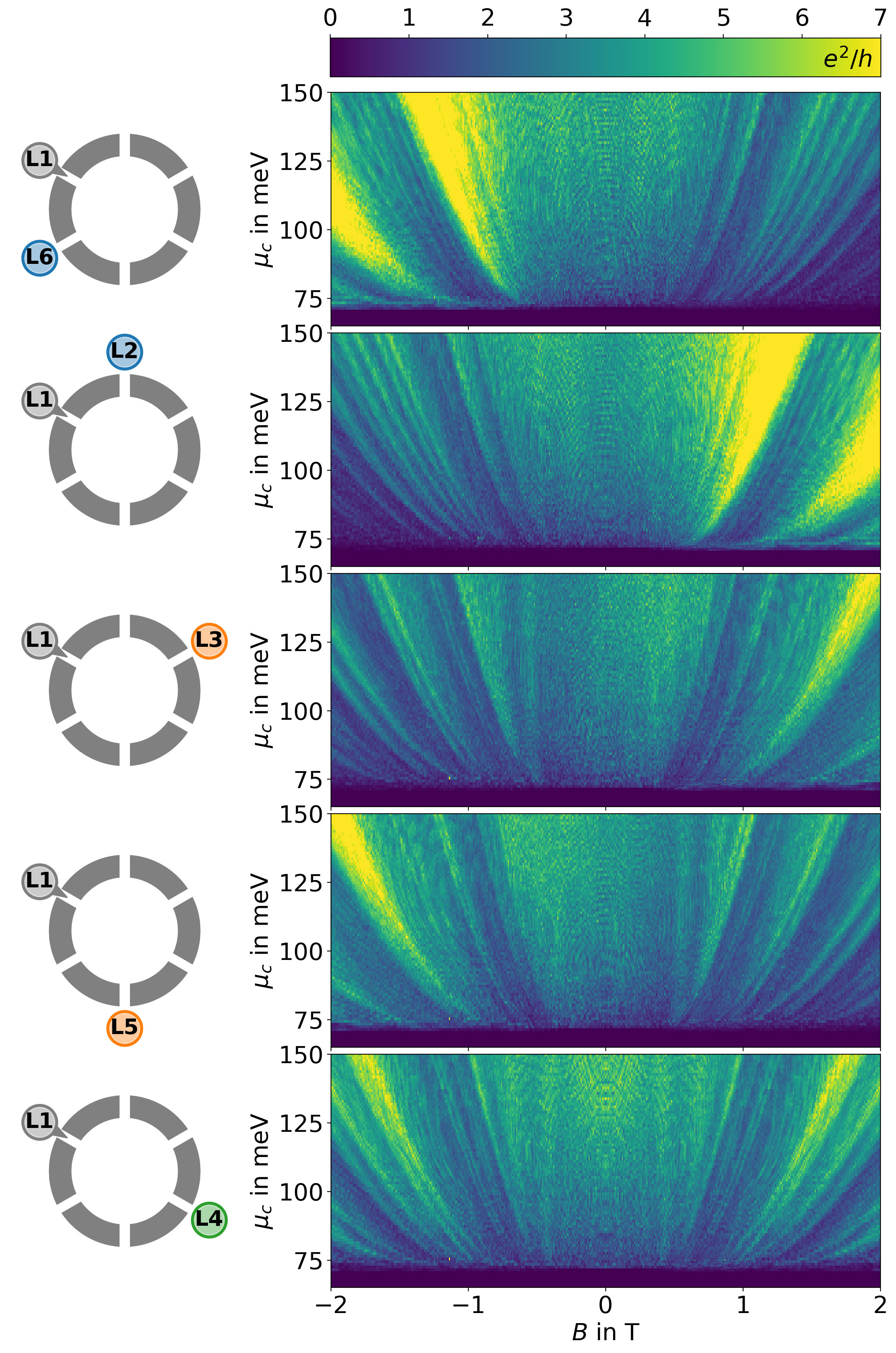}
    \caption{Same as \fig \ref{fig:Magnetotransport}  without trigonally warped Fermi surfaces, i.e., $\gamma_3 = 0$ meV.}
    \label{fig:MagnetoTransportGamma3Off}
\end{figure}

\section{Closed Cavity} \label{App : Closed Cavity}
\begin{figure}
    \centering
    \includegraphics[width=1\linewidth]{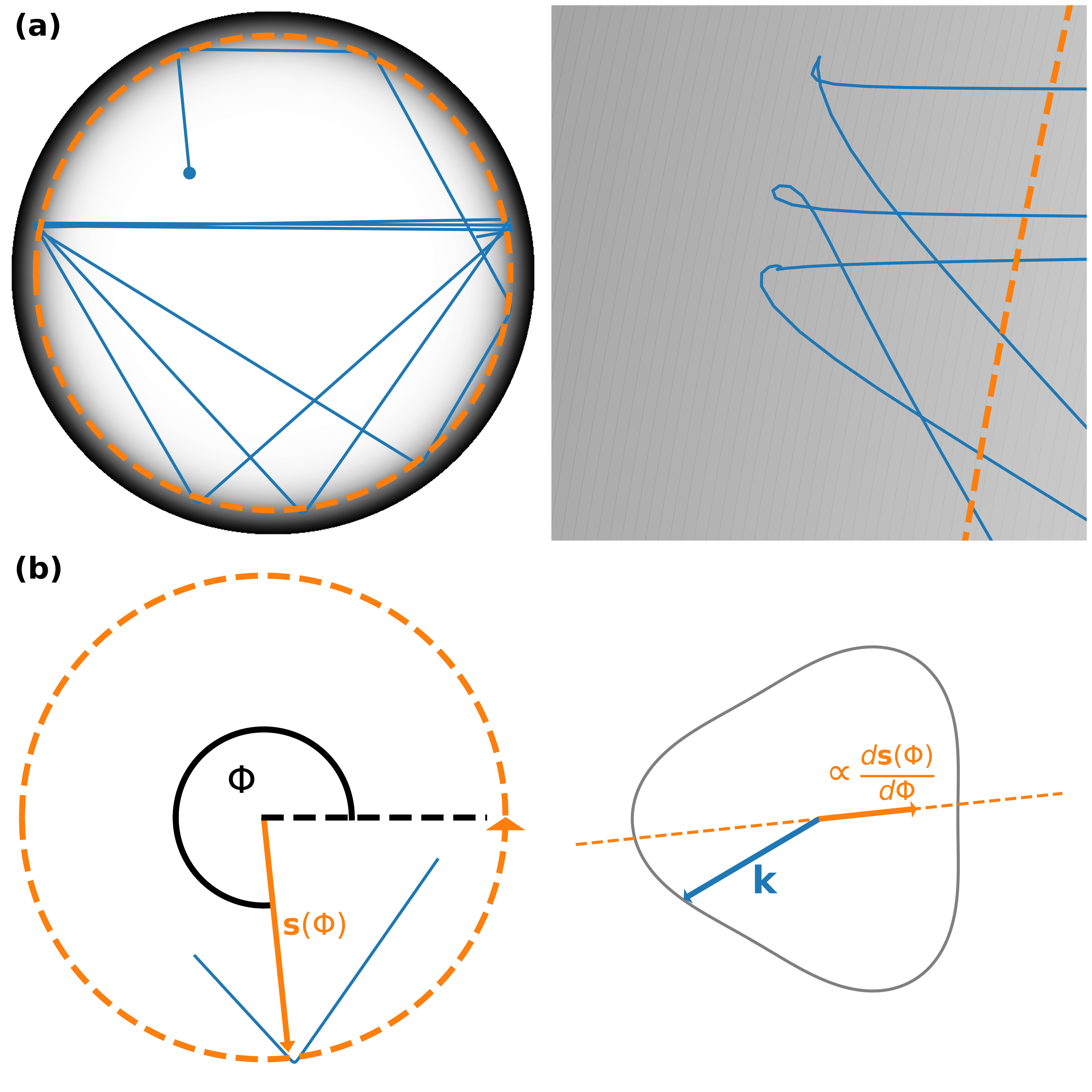}
    \caption{(a) Circular billiard with smooth boundary conditions. In blue, the calculated trajectory of an electron in BLG starting from the blue dot. The orange dashed line depicts the virtual boundary used to measure the parallel part of the incoming momentum. On the left side a close up of the electrons motion inside the boundary region is displayed.
    (b) A schematic depiction of the calculation of the parallel part of the momentum.
    The position of the crossing of the virtual boundary is given by $\mathbf{s}(\phi_c)$.
    The parallel momentum is than obtained by projecting the momentum $\mathbf{k}$ onto the orientation of the virtual boundary at the crossing.}
    \label{fig:ExplainationOfTheBounceMap}
\end{figure}

\begin{figure}
    \centering
    \includegraphics[width=\linewidth]{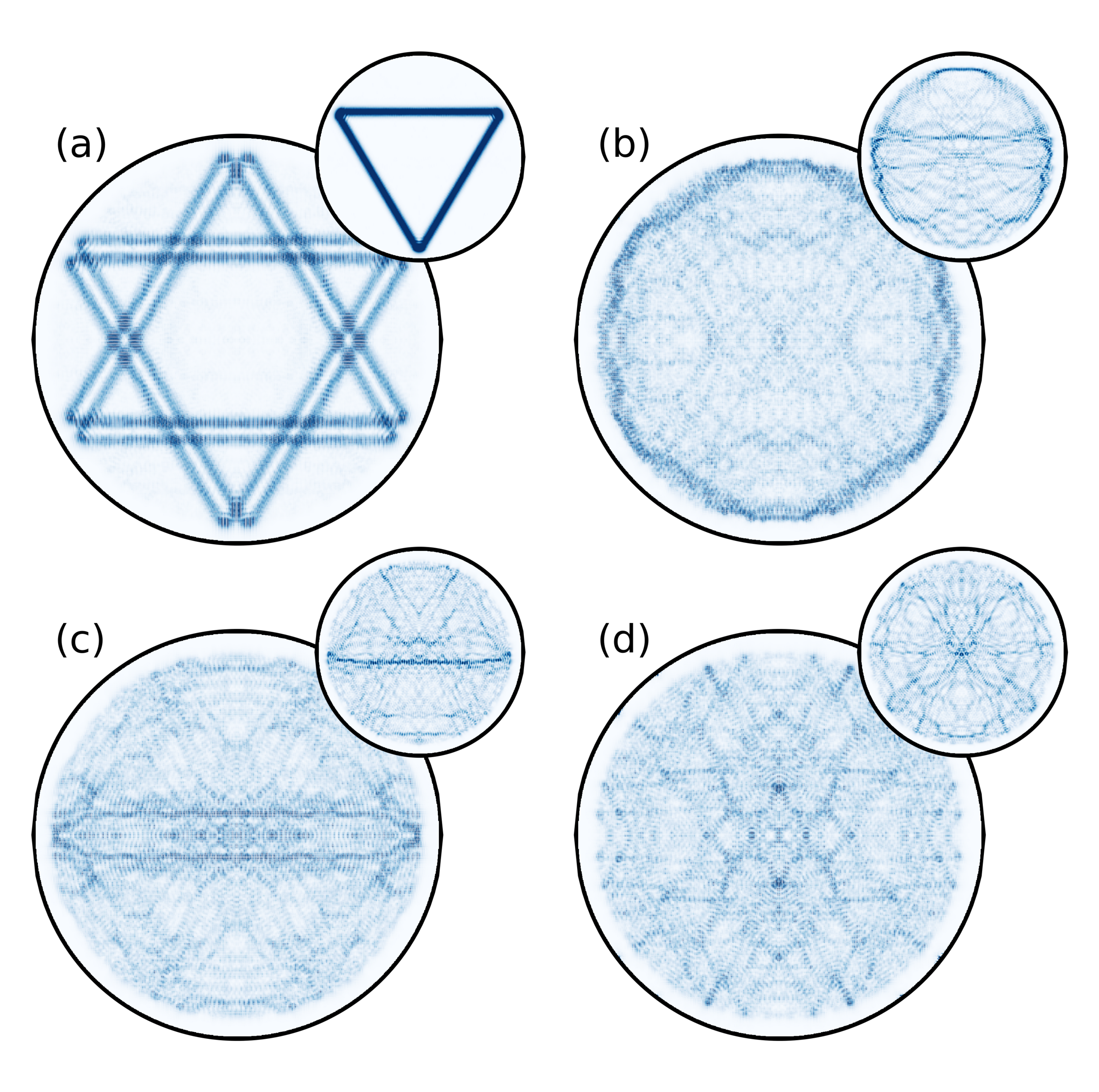}
    \caption{Probability density distributions for tight-binding simulations of the BLG cavity with hard wall boundary conditions. In (b) the gallery modes can be identified. The simulations have been done for $U= 40$ meV and $\mu_c = 100$ meV.}
    \label{fig:HW_LDOS}
\end{figure}

\begin{figure}
    \centering
    \includegraphics[width=\linewidth]{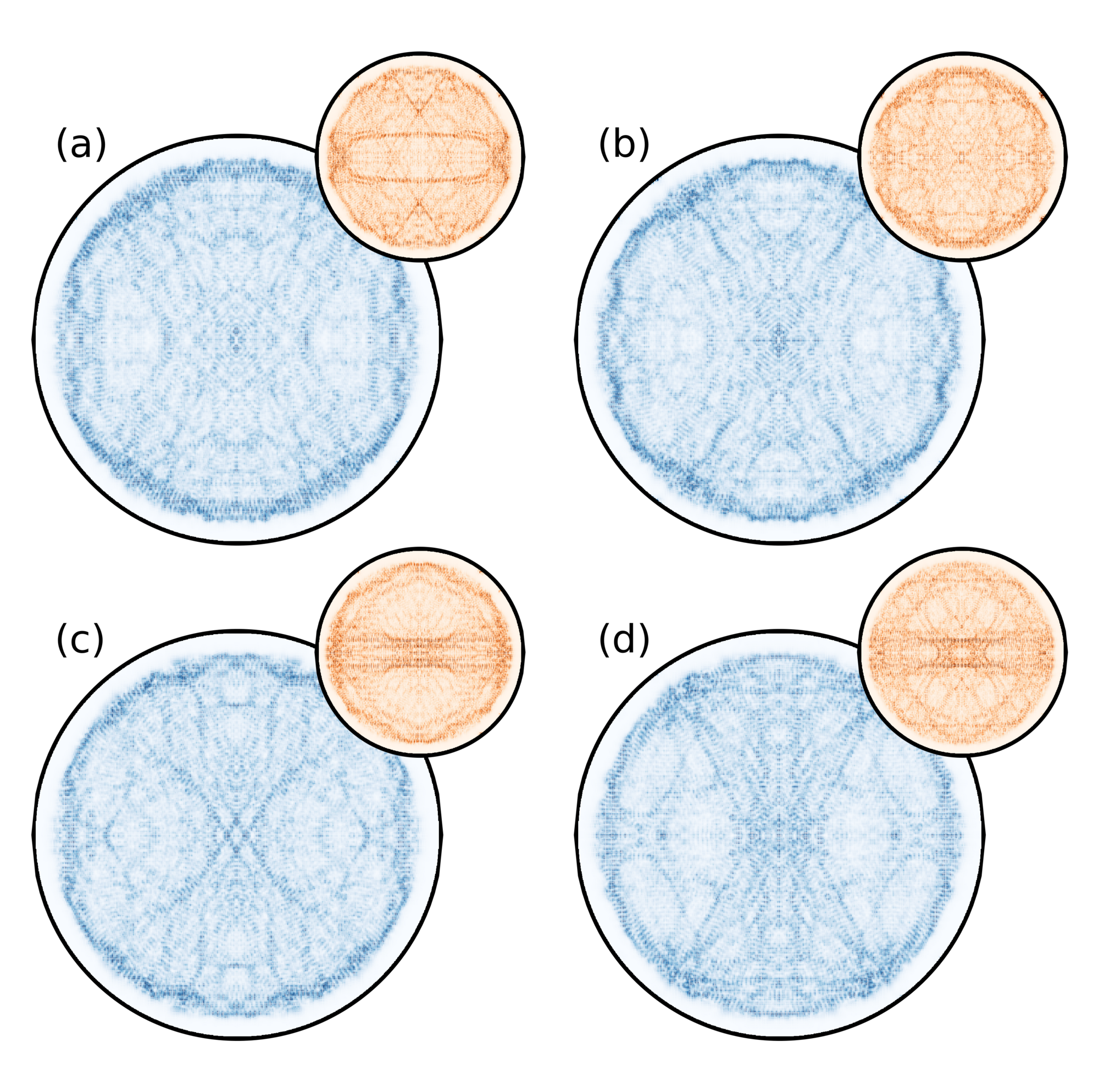}
    \caption{In blue exemplary LDOS for hard wall boundary conditions compared with LDOS for smooth wall boundary conditions (orange in the small insets). Although the chosen LDOS for the smooth boundary exhibit high probability densities at the boundaries of the cavity, they are either found in combination with prominent "bouncing ball" like structures or with gaps at the boundary. For all shown LDOS one should keep in mind that both valleys have been simulated simultaneous and that only half of the gallery like state might correspond to one valley. The calculations have been done for $U= 40$ meV and $\mu_c = 100$ meV.}
    \label{fig:HW_LDOS2}
\end{figure}

We use two approaches to investigate the closed cavities.
First, we change the smooth electrostatic confinement in the tight binding calculations to a hard wall boundary ($d\rightarrow 0$ nm see Eq.\ref{eq:Transition Function}). In this case, we recover the whispering gallery modes found in \cite{seemannGatetunableRegularChaotic2023} (compare \fig \ref{fig:HW_LDOS}) (b) and blue LDOS in \fig \ref{fig:HW_LDOS2}).
As the smoothness of the boundary alone cannot be the reason for the disappearance of the whispering gallery states (since a rotational invariant systems has to support stable whispering gallery states by virtue of its symmetry), we use the semiclassical equation of motions (Eq. \ref{eq:SCL_EOM} and Eq. \ref{eq:SCL_EOM_2}) \cite{neilw.ashcroftFestkorperphysik2001} 
to simulate the trajectories of single electrons inside a circular billiard with smooth electrostatic confinement.
To visualize the system's dynamics, we use a standard version of the Poincar\'e surface of section (PSOS) for billiards,
where we plot the parallel part of the incoming momentum, $k_{||}$,  relative to the boundary, against the position of the particle with respect to the arc-length of the billiard.
As displayed in \fig \ref{fig:ExplainationOfTheBounceMap}(a), inside the smooth boundary region the momentum of the particle changes continuously, making a clear definition of a incoming parallel momentum impossible. 
To avoid this issue we define a virtual boundary $\mathbf{s}$ at the border of the smooth scattering region, for which we calculate the parallel momentum.
The virtual boundary is given by
\begin{align}
    \mathbf{s} = s_0 \left( \begin{array}{c}
         \cos(\phi)  \\
         \sin(\phi)
    \end{array} \right),
\end{align}
where $s_0$ is its radius and $\phi$ the angle measured counterclockwise from the x-axis. 
When the trajectory of the particle crosses the virtual boundary at $\mathbf{s}(\phi_c)$, the parallel momentum of the particle is  given by
\begin{align}
    k_{||} = \frac{1}{s_0} \frac{d\mathbf{s}(\phi)}{d\phi}\Bigg\rvert_{\phi_c} \cdot \mathbf{k}.
\end{align}

\subsection{Breaking the rotational symmetry}
To study the breakdown of the whispering gallery states, we successively increase the perturbation of the rotational symmetry of the system by increasing the triangular warping of the Fermi surfaces.
This we achieve by increasing $\gamma_3$.
In \fig \ref{fig:Bouncemap}, we plot the corresponding PSOS for different values of $\gamma_3$.
For circular Fermi surfaces, the system's rotational invariance demands integrable motion, leading to horizontal structures in the PSOS.
In this case, the whispering gallery states are stable.
Only after considerably large distortion of the Fermi contour $\gamma_3 > 100$ meV the whispering gallery states (high density lines at the top and bottom of the image) vanish.
Combining the facts that in \cite{seemannGatetunableRegularChaotic2023} the whispering gallery states have been found for $\gamma_3 = 381$ meV for hard wall boundary conditions, and that we have found them in tight binding simulations with hard wall boundary conditions, we conclude that both, a triangular warped Fermi surface and a smooth boundary wall are necessary to break those boundary modes.

\subsection{Tuning between integrable and chaotic motion}
\label{subsec:TuningIC}
\begin{figure}
    \centering
    \includegraphics[width=\linewidth]{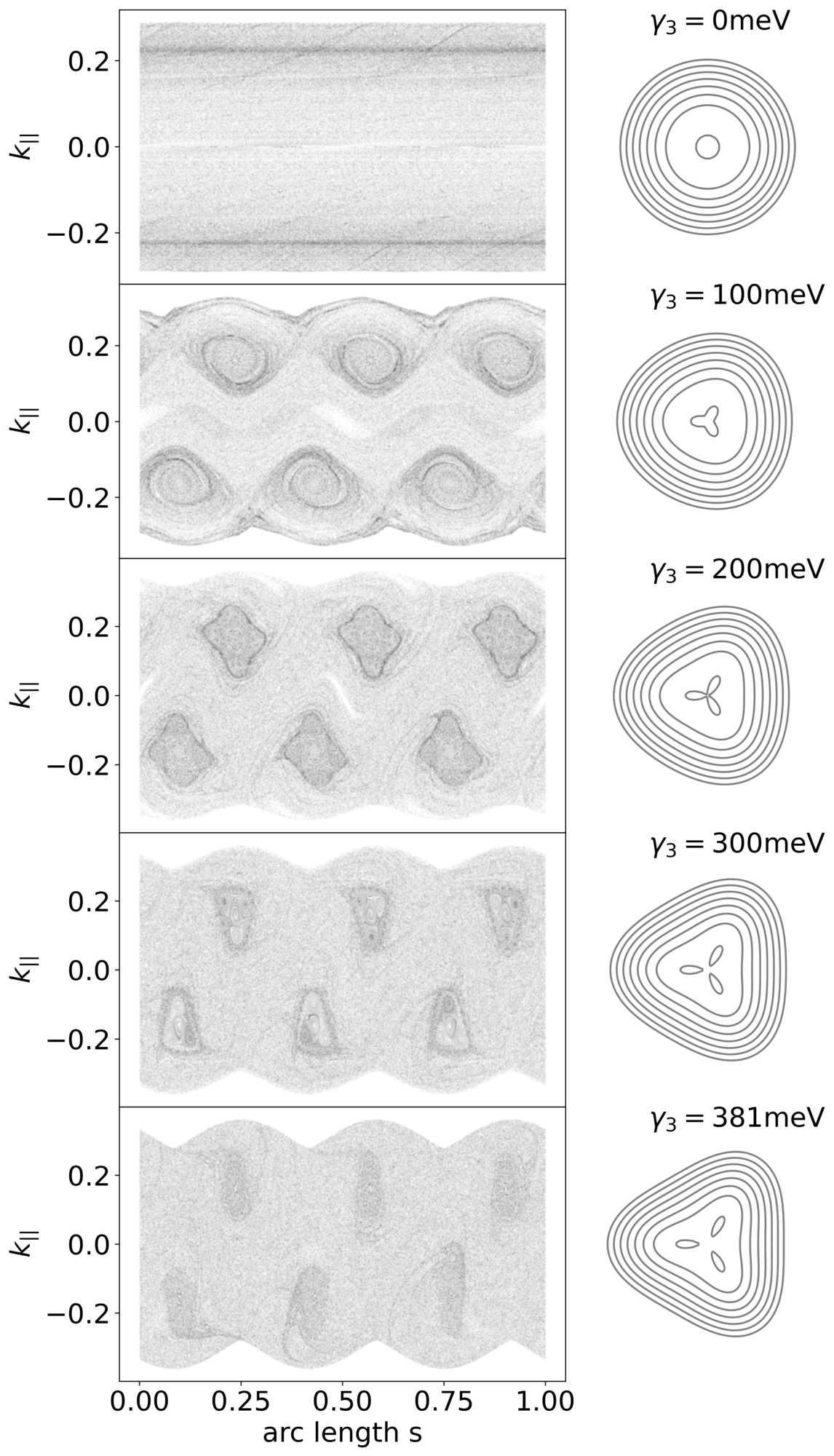}
    \caption{PSOS for a circular billiard with smooth boundary conditions.
    From top to bottom $\gamma_3$ and therefore the trigonal warping of the Fermi contours is increased (displayed to the right). The displacement field is $U = 40$ meV and $\mu_c= 100$ meV.}
    \label{fig:Bouncemap}
\end{figure}

\begin{figure}
    \centering
    \includegraphics[width=\linewidth]{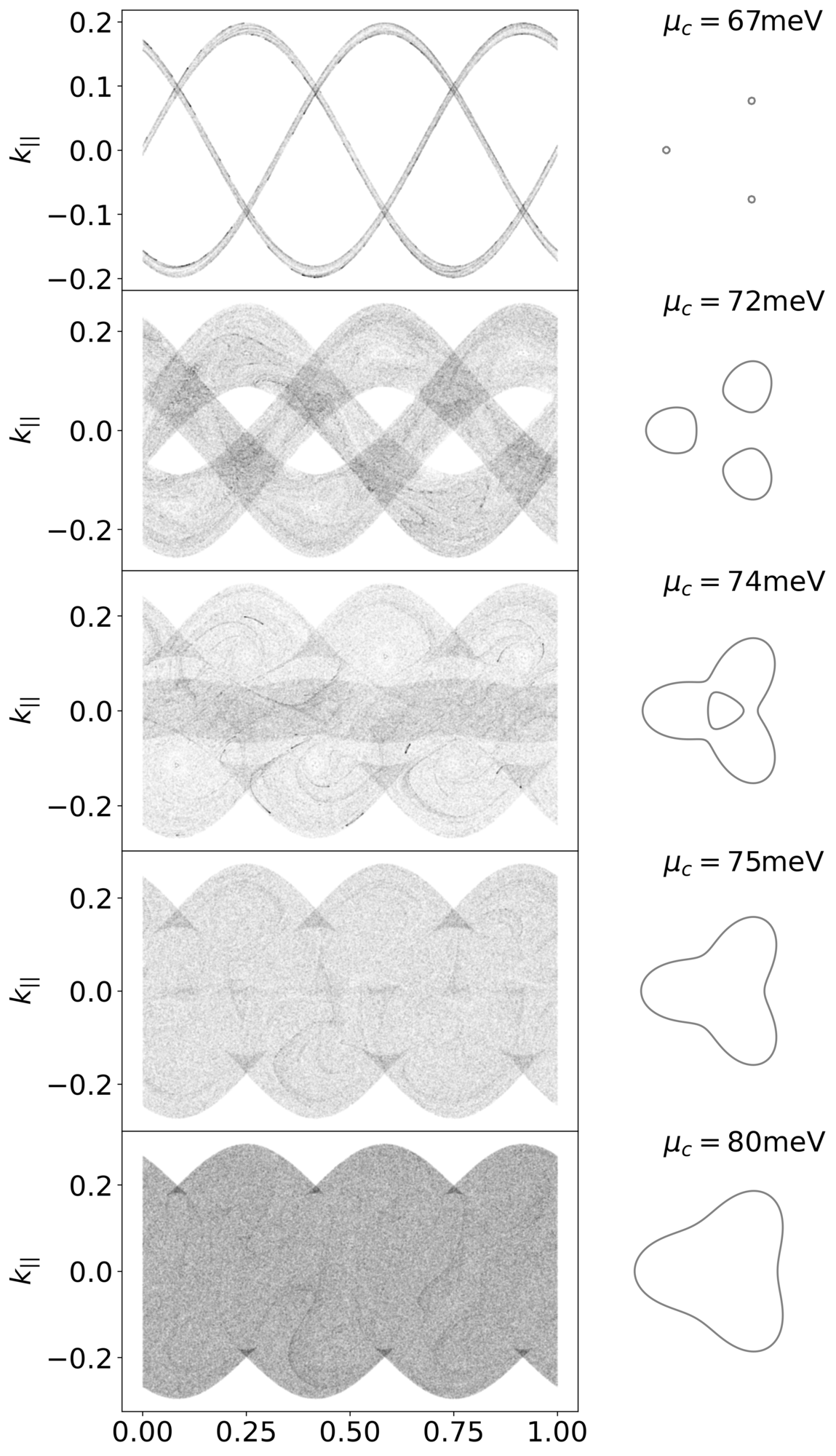}
    \caption{PSOS for different Fermi energies $\mu_c$ and a fix displacement field of $U=150$ meV. The Lifshitz transition allows to tune between regular motion ($\mu_c = 67$ meV) and chaotic motion ($\mu_c = 80$ meV).}
    \label{fig:PoincareLifshitz}
\end{figure}

While a change of the $\gamma_3$ parameter is a purely academic exercise, the Lifshitz transition of the system allows to tune between integrable and chaotic motion inside a BLG.
To this end, we show in \fig \ref{fig:PoincareLifshitz} the PSOS for different potential strengths $\mu_c$ in the case of a large distortion field of $U=150$ meV.
For small potential strengths, the three Fermi surfaces denoting the mini-valleys are circular and lead in combination with the circular billiard to integrable motion.
In contrast to the integrable motion depicted by horizontal lines as seen in \fig \ref{fig:PoincareLifshitz}, one finds a sinusoidal structure in the PSOS, which we will briefly discuss in the following.
For circular Fermi contours, one can parameterize the momentum by
\begin{align}
    \mathbf{k}(\beta) = k_0 \left( \begin{array}{c}
         \cos(\beta)\\
         \sin(\beta) 
    \end{array} \right),
\end{align}
where $k_0$ is the absolute value of the momentum and $\beta$ the angle dictating the orientation of the momentum vector in k-space.
In the case of circular Fermi contours, the direction of the group velocity and the direction of the momentum vector are identical.
For a hard wall boundary condition the angles for consecutive scattering events are governed by the standard map \cite{tabachnikovGeometryBilliards2005}
\begin{align}
    \phi_n = \phi_0 +(\pi -2 \alpha) n \\
    \beta_n = \beta_0 + (\pi -2\alpha) n,
\end{align}
where $\phi_0,\beta_0$ are the initial angles of a trajectory.
Here $\alpha$ is the incident angle - which is conserved for circular billiards in combination with circular Fermi surfaces - and $n$ the number of occurred scattering events.
As a result, the parallel momentum becomes
\begin{align}
    k_{||} = \frac{1}{s}\frac{d \mathbf{s}(\phi)}{d\phi} \cdot \mathbf{k} =k_0 \sin(\beta_0 -\phi_0),
\end{align}
which results in horizontal structures in the PSOS (compare \fig \ref{fig:Bouncemap}).
When the center of the $k$ vector is shifted from the origin by a constant vector $\mathbf{k}_{\text{shift}}$ - as it is the case for the three mini-valleys - the parallel part of the momentum vector becomes 
\begin{align*}
    k_{||} &= \frac{1}{s}\frac{d \mathbf{s}(\phi)}{d\phi} \cdot (\mathbf{k}+\mathbf{k}_{\text{shift}}) \\
    =k_0 \sin(\beta_0 -\phi_0) &- \mathbf{k}_{\text{shift}} \cdot \left(  \begin{array}{c}
         - \sin(\phi_0 + (\pi-2\alpha)n)  \\
         \cos(\phi_0 + (\pi-2\alpha)n) 
    \end{array}          \right),
\end{align*}
and the oscillating behavior emerges.

Adding the motions corresponding to the three mini-valleys together, one obtains the braid like structure.
For regular motion, each braid is clearly structured by parallel non crossing sinusoidal lines.

With increasing energy, the braids broaden, as $k_0$ increases and the internal structure of the braids breaks due to the deformation of the mini-valleys.
For high enough energies, the Lifshitz transition occurs and the braids merge together forming the  familiar PSOS of the singly connected, trigonal warped BLG Fermi line.

\clearpage
\bibliography{Lib_ud}
\end{document}